# Optically-Controlled Nano-Transducers Based on Cleaved Superlattices for Monitoring Gigahertz Surface Acoustic Vibrations

Changxiu Li[1], Nikolay Chigarev[1], Théo Thréard[1], Kedong Zhang[2], Nicolas Delorme[3], Vincent Tournat[1], Samuel Raetz[1], Hong Lu[2], Vitalyi E. Gusev[1,*]

[1]Laboratoire d'Acoustique de l'Université du Mans (LAUM), UMR 6613, Institut d'Acoustique – Graduate School (IA-GS), CNRS, Le Mans Université, Le Mans, France
[2]College of Engineering and Applied Sciences, Nanjing University, Nanjing, China
[3]Institut des Molécules et Matériaux du Mans (IMMM), UMR 6283 CNRS, Le Mans Université, Le Mans, France

* vitali.goussev@univ-lemans.fr

**ABSTRACT**

Surface acoustic waves (SAWs) convey energy at subwavelength depths along surfaces. Using interdigital transducers (IDTs) and opto-acousto-optic transducers (OAOTs), researchers have harnessed coherent SAWs with nanosecond periods and micrometer localization depth for various applications. These applications include the sensing of small amount of materials deposited on surfaces, assessing surface roughness and defects, signal processing, light manipulation, charge carrier and exciton transportation, and the study of fundamental interactions with thermal phonons, photons, magnons, and more. However, the utilization of cutting-edge OAOTs produced through surface nanopatterning techniques has set the upper limit for coherent SAW frequencies below 100 GHz, constrained by factors such as the quality and pitch of the surface nanopattern, not to mention the electronic bandwidth limitations of the IDTs. In this context, unconventional optically-controlled nano-transducers based on cleaved superlattices (SLs) are here presented as an alternative solution. To demonstrate their viability, we conducted proof-of-concept experiments using ultrafast lasers in a pump-probe configuration on SLs made of alternating $Al_xGa_{1-x}As$ and $Al_yGa_{1-y}As$ layers with approximately 70 nm periodicity and cleaved along their growth direction to produce a periodic nanostructured surface. The acoustic vibrations, generated and detected by laser beams incident on the cleaved surface, span a range from 40 GHz to 70 GHz, corresponding to the generalized surface Rayleigh mode and bulk modes within the dispersion relation. This exploration shows that, in addition to SAWs, cleaved SLs offer the potential to observe surface-skimming longitudinal and transverse acoustic waves at GHz frequencies. This proof-of-concept demonstration below 100 GHz in nanoacoustics using such an unconventional platform offers opportunities for realizing sub-THz to THz coherent surface acoustic vibrations in the future, as SLs can be epitaxially grown with atomic-scale layer width and quality.

Surface acoustic waves (SAWs) possess the distinct ability to interact with various state variables within solids. These variables include electrons, spins, photons, magnons, thermal phonons, and strain, enabling the evaluation and/or modification of material properties at subwavelength depths near the surface. The significance of SAWs spans over six decades, beginning with the invention of interdigital transducers (IDTs) in 1965[1] and their subsequent use in combination with lasers for SAW monitoring in 1968.[2] While lower-frequency SAWs, falling within the ultrasonic range, have been influential in communication and signal processing technologies,[3] higher-frequency SAWs have found applications in diverse fields. These application include material characterization,[4] photonic modulation,[5,6] optomechanics,[7] the transport of other excitations,[8-10] and interactions with magnons.[11] SAWs are involved in a variety of products for Information and Communication Technologies (ICTs), including mobile devices (utilized in central processing units, signal converters and gyroscopes) and devices associated with the "Internet of Things", where they find applications in sensors and integrated energy converters.

At present, the upper operational frequency limit for coherent SAWs controlled by IDTs or opto-acousto-optic transducers (OAOTs) remains below 100 GHz.[12,13] Extending this frequency range to sub-terahertz (sub-THz) range in integrated elasto-optic modulators and SAW processors/filters holds significant potential for future ICT devices. Access to higher frequencies and broader channel bandwidths is considered essential for the continued expansion of global mobile communications. Sub-THz SAWs could provide the means to characterize materials at depths in the single-digit nanometer range beneath the surface. They could be also applied to characterize materials or molecules deposited between the SAW emitter and detector along the path of SAW propagation. A primary objective of an ongoing research is to develop techniques for exciting and detecting coherent SAWs with picosecond periods and deeply sub-optical, nanometer-scale wavelengths and localization depths for applications in nanometrology, nanoimaging, as well as in sensing and manipulations for both fundamental and applied research, and the field of ICTs.

From a theoretical perspective, the generation of SAWs at a specific frequency necessitates the creation of a spatio-temporal stress distribution in a material at the SAW penetration depth during excitation. The temporal Fourier spectrum of this distribution should include the desired frequency component. Simultaneously, its spatial Fourier spectrum along the surface should encompass the required SAW wavenumbers.[14] The expansion of IDTs to sub-THz frequencies faces limitations due to electronics bandwidth, unlike laser-based methods, which lack such restrictions. Intensity envelopes of laser pulses lasting less than a picosecond contain frequencies up to the terahertz range, effectively incorporating these frequencies into the spectrum of photo-induced stress.

Creating the required high wavenumbers in the spatial distribution of photo-induced stress can be accomplished using laser gratings. This technique involves the interference between two overlapping laser beams propagating at an angle, forming an intensity pattern with periodicity that selects the SAW wavenumber.[15-17] In this technique, the ratio of the wavevector of SAW to that of electromagnetic radiation is less or equal to 2.[17,18] Notably, in laser methods, SAWs are detected by observing the scattering of probe laser radiation due to SAW-induced changes in the material refractive index and/or the surface topography. The highest-frequency Rayleigh SAW (50 GHz) was generated and detected using extreme ultraviolet (EUV) light grating with an 84 nm period.[18] Femtosecond pump pulses at a wavelength of 39.9 nm and a probe pulse at a wavelength of 13.3 nm were derived from the FERMI free electron laser source. EUV radiation from the free electron laser was also employed to induce gratings with a minimum period of 28 nm, although this allowed monitoring of only the longitudinal acoustic waves propagating along the grating direction.[19] Due to the very high velocities of the longitudinal waves in the sample, the 28 nm EUV gratings provided the capability to monitor longitudinal acoustic waves up to 360 GHz in frequency.

As an alternative, the nanometer spatial periodicity of laser-induced stresses can be achieved through surface nanostructuring. Historically, material nanopatterning has been accomplished through depositing nanoscale arrays of metal lines (Figure 1a). In these cases, SAWs were predominantly generated via the thermoelastic stress induced by laser heating of such grating.[20,21] Experiments with metallic gratings having a minimum period of around 40 nm were conducted using 800 nm infrared pump laser pulses and 30 nm EUV probe laser pulses.[22] The highest reported coherent SAW frequency was approximately 90 GHz, achieved in 100 nm period metallic gratings pumped and probed using near infrared light (at around 820 nm wavelengths),[13] indicating successful generation and detection with deeply sub-optical gratings.

However, for SAWs to reach 1 THz, the grating period controlling the wavenumber should be shorter than 3 nm, assuming a typical Rayleigh SAW velocity of 3000 m/s in the material. The minimum period of structures currently patterned in leading research laboratories is around 10 nm.[23-25] Therefore, the primary limitation in generating SAWs at frequencies near 1 THz is the need to reduce surface nanopatterning pitch while improving its quality. We propose that this limitation can be addressed by engineering unconventional OAOTs based on cleaved nanostructured bulk materials. Bulk superlattices (SLs) with a period of several atomic layers and atomic-scale quality at the interfaces can currently be grown by epitaxy. These SLs can be cleaved or sliced (*i.e.*, with microtome) along the surface normal to their layering, producing a nanopatterned surface, with the periodicity and quality of bulk samples, that can control SAWs (Figure 1b). Figure 1c depicts the consecutive steps of a cleaved SL attainment from a bulk SL grown on a substrate through cleavage. While the highest frequencies of coherent bulk acoustic waves up to 3 THz have been generated and detected by lasers in SLs/multiple quantum well structures,[26,27] they have not yet been applied to control SAWs. Previously, the cleaved SL structure was theoretically investigated.[28] This structure is stratified normal to the cleaved surface and has physical parameters varying periodically not only in the bulk but also at its cleaved mechanically free surface. It was studied as a SAW phononic crystal, and an increased efficiency of the optoacoustic (OA) conversion in cleaved semiconductor SL for monitoring GHz SAWs was suggested.[29,30] Note that we will henceforth refer to the experimental SL structure with notation "SL stratified normal to the surface", as suggested in the pioneering theoretical investigation.[28] While the theoretical investigations were conducted more than three decades ago, experimental demonstrations of acoustic waves on such nanostructures have not yet been realized. Here, on the platform of cleaved semiconductor SLs, we conducted proof-of-concept experiments for all-optical monitoring of coherent acoustic vibrations of the surface (Figure 1b), encompassing various acoustic modes, including generalized Rayleigh-type (gR) surface acoustic waves, surface-skimming transversal (T), and longitudinal (L) acoustic waves, all of which are reported within this work. Two samples have been tested. The first, an GaAs/AlAs SL with a 71 nm period, allowed us to simultaneously monitor the gR, T, and L waves at frequencies of 40.2 GHz, 50.9 GHz, and 70.3 GHz, respectively, using near-infrared femtosecond lasers. In the second sample, containing two $Al_{0.2}Ga_{0.8}As/Al_{0.4}Ga_{0.6}As$ SLs of the same period but with a modified content of Al in constituent layers compared to the first sample, the gR wave at 40 GHz frequencies was also monitored.

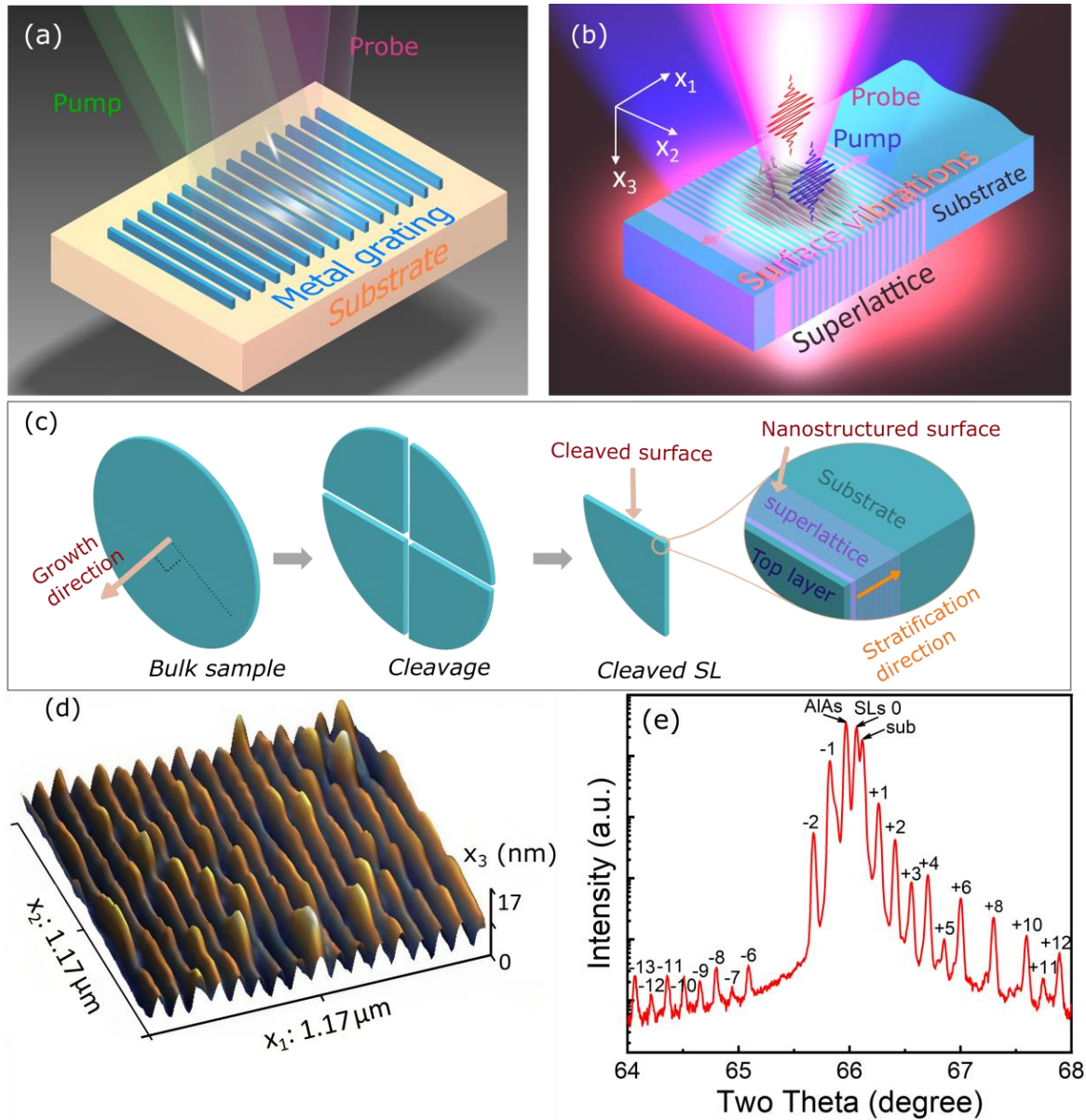

Figure 1. (a) An illustrative display of a metal grating structure. (b) Schematic of pump-probe experiments on the cleaved SL. (c) The illustration of the experimental sample preparation. GaAs substrate with the grown on it SL and cap GaAs layer is denoted as bulk sample. (d) AFM image of a cleaved GaAs/AlAs SL surface modified by oxidation. (e) X-ray diffraction measurement of a GaAs/AlAs SL.

**RESULTS AND DISCUSSION**

The nanostructures were prepared by epitaxial growth along the (001) direction of a semiconductor SL on a GaAs substrate, layer by layer, with precise control of each layer composition (alternating between $Al_xGa_{1-x}As$ and $Al_yGa_{1-y}As$) and thickness (alternating between $d_1$ and $d_2$). This growth resulted in a periodic stack of nanometers-thin layers stratified normal to the substrate surface. Two semiconductor structures were investigated. They contain SLs with a nominally equal period and individual layer thickness of $d_{SL}$ = 71.4 nm, $d_1$ = 40.0 nm, $d_2$ = 31.4 nm, while the constituent materials are GaAs/AlAs and $Al_{0.2}Ga_{0.8}As/Al_{0.4}Ga_{0.6}As$, respectively. The first sample contains a single SL of 5.35 μm thickness, and the second one contains two SLs of 1.2 μm thickness separated by a 1 μm-thick GaAs layer. Cleavage procedures were applied to the sample along the growth direction, perpendicular to the substrate surface, producing a mirror-finished nanostructured surface for optical monitoring of coherent acoustic phonons (Figure 1b).

However, oxidation of $Al_xGa_{1-x}As$ (x ≠ 0) layers exposed to air occurred. The morphology of the nanostructured surface two months after cleavage was revealed by atomic force microscopy (AFM, see Methods A), showing ripples with a periodicity of 71nm, extracted from a 2D FFT analysis of the image. Apart from local defects, the average modulation depth of the oxidation induced ripples is 7 nm in the GaAs/AlAs SL (see Figure 1d). Therefore, the sample with the nanostructured surface (Figure 1c) was freshly cleaved just before the pump-probe experiments to minimize oxidation effects, although complete prevention was not achieved.

It is worth noting that in the second sample, the Al content is sufficiently low in both SL layers to significantly reduce the surface oxidation effect.[31] The X-ray diffraction (XRD) results on the GaAs/AlAs SL confirmed the good quality and periodicity of the structure (see Figure 1e and Methods A). In contrast to a metal grating deposited on a substrate (see Figure 1a), the cleaved SL structures (Figure 1b) acquire their nanopatterned surface through the nanostructuring of bulk SLs. Common techniques such as lithography and focused ion beam are employed for nanostructuring only the surface.

The periodicity of the mechanical/acoustical properties (such as density and elastic moduli) in the cleaved SLs gives rise to Brillouin zone-center acoustic eigenmodes with corresponding lateral, *i.e.*, along $x_1$ direction on the cleaved surface, periodicity of the mechanical strain/displacement fields. These modes, for example, gR waves, contain lateral unmodulated/homogeneous components, which correspond to infinitely long modulation period (see Methods B). The same lateral periodicity of the optical properties, *i.e.*, of the complex optical refractive index, significantly enhances the possibilities for optically generating these acoustic eigenmodes (see Methods C). The identical periodicity of both optical and acousto-optical properties, *i.e.*, of the complex photoelastic tensor, facilitates the optical detection of these modes (see Methods D).

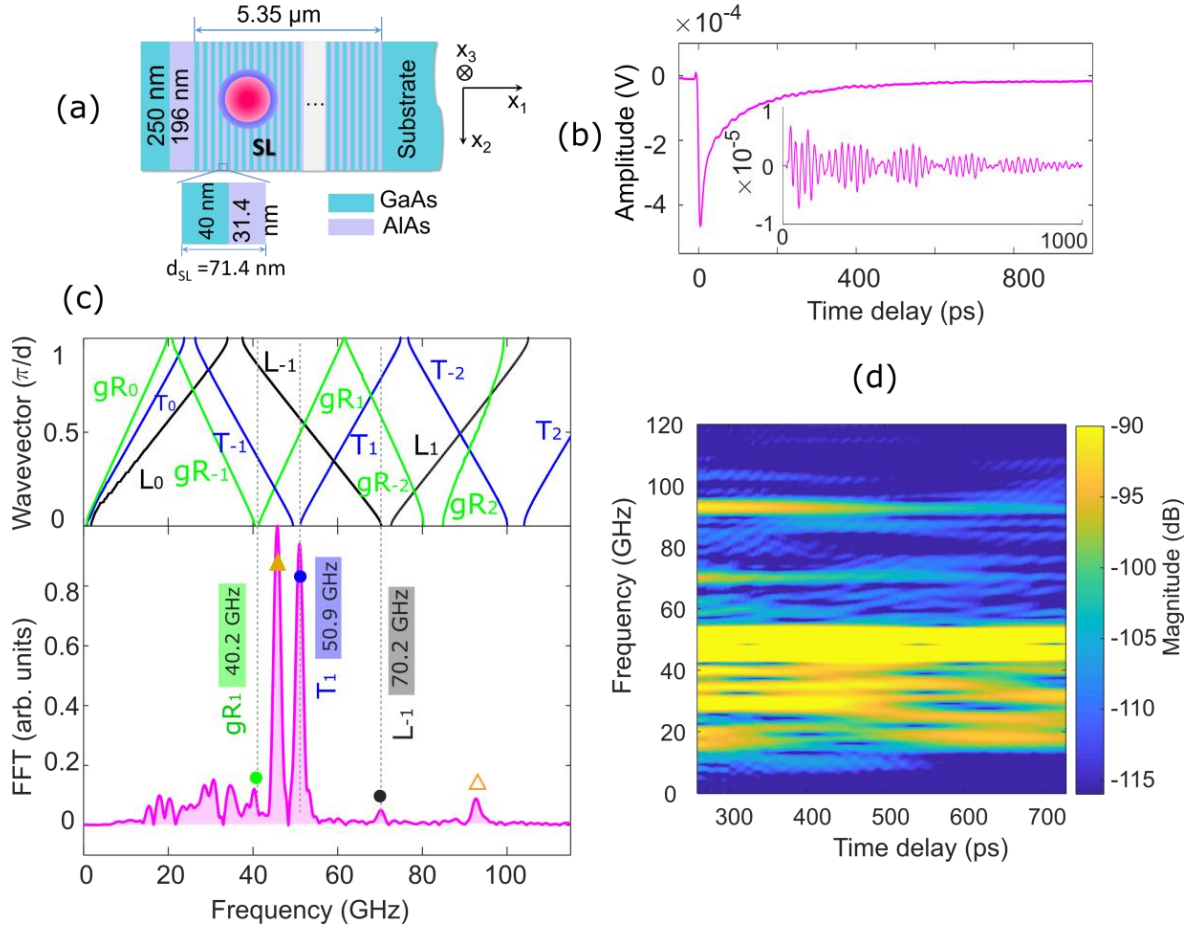

Figure 2. Presentation of the scheme and of evaluation results for the first sample. (a) Schematic depiction of the sample surface and of a potential laser focus position. (b) Time-domain reflectivity signal. Inset: acoustically-induced oscillations in the transient reflectivity signal after removing non-oscillating background. (c) Upper part: calculated dispersion relations of acoustic modes in the SL, which is infinitely thick along $x_1$ and semi-infinite along $x_3$ (green: gR, blue: T, black: L). Subscripts indicate acoustic band orders; lower part: spectrum of the acoustic vibrations presented in the inset of (b). Green, blue and black solid dots indicate the identified first-order $k_1 = 0$ gR, T, and L waves, respectively. Brown solid and hollow triangles represent the frequencies of bulk longitudinal acoustic waves propagating in the zeroth (45.7 GHz) and first (92.6 GHz) diffraction orders, respectively. (d) Spectrogram of the acoustic signal shown in the inset of part (b) obtained by short-time Fourier transform (see Methods F for details).

It is important to note that the existence of laterally unmodulated and homogeneous eigenmode components is theoretically advantageous for their all-optical monitoring. High-frequency acoustic modes at the zone center can be excited even by the laterally homogeneous part of the distributed photo-induced stresses acting on laterally homogeneous components in the mode structure. These modes can be detected through time-domain backward Brillouin scattering of the normally incident probe light, even when the SL motion is laterally unmodulated, meaning that scattering occurs from the motion of the SL averaged over its period (see Methods C and D). Due to the short periodicity in their physical properties, the designed SLs stratified normal to the surface ($x_1x_2$ plane in Figure 1b)[28] offer the opportunity to optically monitor surface acoustic vibrations with deeply sub-optical lateral periodicity, since they are composed of the acoustic waves with deeply sub-optical lengths.

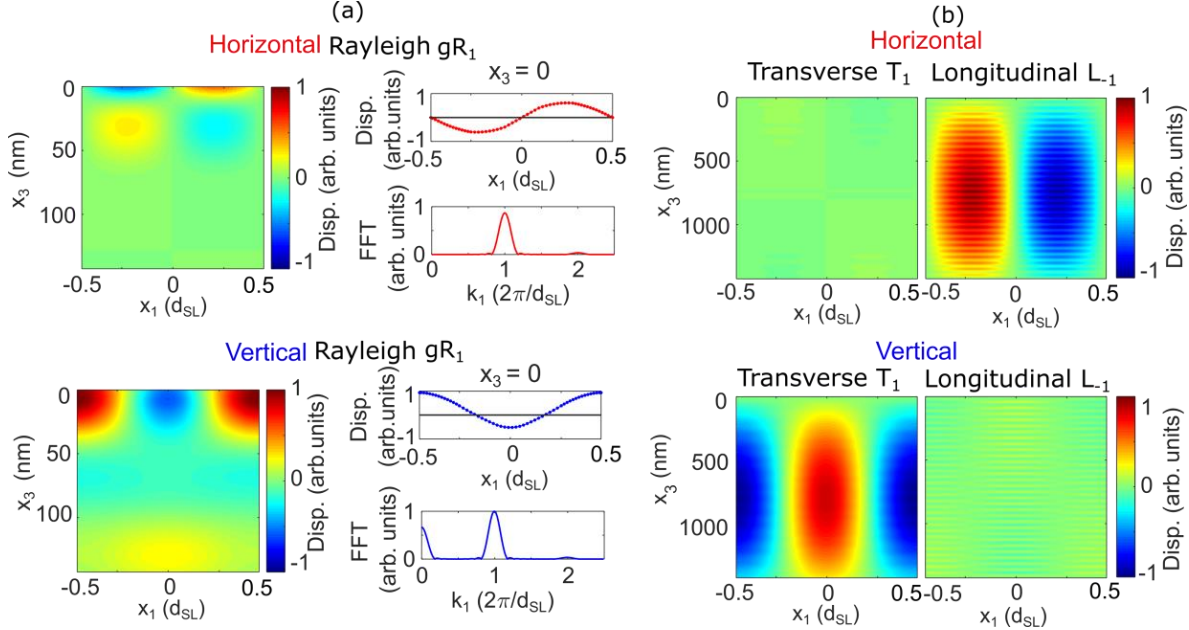

Figure 3. Calculated spatial distributions of the mechanical displacement components in the first-order zone center $k_1 = 0$ symmetric (a) gR (41.4 GHz), (b) T (51.2 GHz) and L (70.3 GHz) acoustic modes. Disp. denotes displacement. The second column in (a) presents the displacements at mechanically free surface and their $k_1$ spectra, highlighting the presence of the spatially homogeneous component in the gR mode structure.

We conducted ultrafast pump-probe transient reflectivity experiments to monitor coherent acoustic phonons on the cleaved SL surface of the prepared samples (see Figure 1b and Methods A). Unlike the conventional setup for monitoring bulk acoustic waves in SLs,[26] where the pump and the probe laser beams are incident on the plane $x_2x_3$, in this case, they are directed normally to the cleaved plane $x_1x_2$. Laser wavelengths of 405 nm/810 nm were chosen for the pump and probe laser beams, which were collinearly co-focused by a 100× objective onto the sample, creating a spot with a radius of approximately 0.82 μm (measured at $1/e^2$ by cross-correlation, see Methods E for details). We monitored the time-dependent reflectivity change of the sample, induced by the pump excitation on the cleaved structure, using the probe light whose intensity was modulated by this change, and we captured it with a photodetector.

The experimental results obtained from both SL-based structures are now presented and compared favourably with theoretical expectations. The proof-of-concept experiment was conducted on the first sample, which was based on the cleaved 5.35 μm-thick GaAs/AlAs SL. In this configuration, the laser beams were focused entirely on the SL region (see Figure 2a). The interband absorption of the pump laser pulses in the GaAs creates electron-hole pairs, whose distribution dominantly contributes to the generation of acoustic waves through the stress induced by the electron-hole-phonon deformation potential mechanism.[32] Notably, this opto-acoustic conversion mechanism is much more efficient than the thermo-elastic one, most commonly found in metallic grating OA transducers.[13,14] The reflectivity change induced by the acoustic waves (phonons) is monitored by the near infrared probe light as a function of the time-delay between the pump and the probe pulses.

The time-domain signal (Figure 2b, main part) reveals that the acoustically-induced oscillations (Figure 2b, inset) are superimposed on the monotonous exponential variations in amplitude caused by the generation/recombination of the electron-hole pairs and transient heating of the sample after photo-excitation. To enhance the display of acoustic oscillations (Figure 2b, inset), the dominant background is numerically removed, following the signal processing details in Methods F. Due to the SL periodicity, the projections $k_1$ of the wave vectors on $x_1$ axis, which are larger than $\pi/d_{SL}$ in the acoustic modes dispersion relations, are folded into a mini-Brillouin-zone (see Figure 2c, upper section, and Methods G for details on numerical calculations). In this zone, the gR, T, and L waves differ in frequencies due to the differences in their propagation velocities. The gR waves become supersonic above the frequency corresponding to the intersection of the first order optical surface phonon branch $R_{-1}$ and the bulk acoustical phonon branch $T_0$, leading to attenuation through the emission of bulk acoustic waves (Brekhovskich attenuation).[33]

However, due to the small relative differences between the parameters of the SL constituent materials, $\mu \ll 1$, where $\mu$ represents the ratios of parameter differences to average parameter values, such as $\Delta C_{11}/\langle C_{11}\rangle = 0.0117$, $\Delta C_{12}/\langle C_{12}\rangle = 0.058$, $\Delta C_{44}/\langle C_{44}\rangle = 0.00845$ and $\Delta\rho/\langle\rho\rangle = 0.3368$ for GaAs/AlAs, a basic analysis (see Supporting Information S1) demonstrates that the $k_1 = 0$ gR wave experiences a slight frequency shift and a weak attenuation proportional to $\mu^2$ ($\Delta\text{Re}(\omega_{gR})/\omega_{gR} \sim \mu^2$, $\text{Im}(\omega_{gR})/\omega_{gR} \sim \mu^2$, with $\omega$ denoting the cyclic frequency). The numerical simulations of the eigenmode dispersion relations in the laterally-infinite cleaved SL shown in the upper panel of Figure 2c, enable the identification of three peaks. These peaks, found in the experimental spectrum in the lower section of Figure 2c at frequencies 40.2 GHz, 50.9 GHz and 70.2 GHz, correspond to the zone center vibrational motions induced by the gR, T, and L waves, respectively. Figure 2d depicts the spectrogram of the acoustic signal in Figure 2b and shows how the frequency contents of the signal vary with time delay. This will be further discussed later on.

Figure 3 illustrates the calculated shapes of these three modes, revealing that the vibrations caused by the gR wave (Figure 3a) are highly localized near the surface (at depths smaller than one SL period). In contrast, the vibrations induced by the T and L waves (Figure 3b) are non-localized. To better appreciate the surface displacement at $x_3 = 0$ induced by the gR wave, the horizontal (top) and vertical (bottom) displacement are displayed in the second column of Figure 3a as a function of $x_1$. It is important to note here that both components depict spatial periodicity associated with the SL periodicity $d_{SL}$. More importantly, the vertical component, unlike the horizontal one, contains a non-zero mean value as seen from the non-zero surface area and, related non-zero Fourier component at $k_1 = 0$ of its displacement distribution. The differences in amplitude of the related transient reflectivity oscillations between the gR and T waves at closely matched frequencies can likely be attributed to the oxidation of AlAs layers and the imperfect cleavage procedures, which introduce surface defects and roughness (as shown in Figure 1d). These factors are expected to exert a more significant influence on the generation, propagation, and detection of surface-localized acoustic eigenmodes compared to their delocalized counterparts.

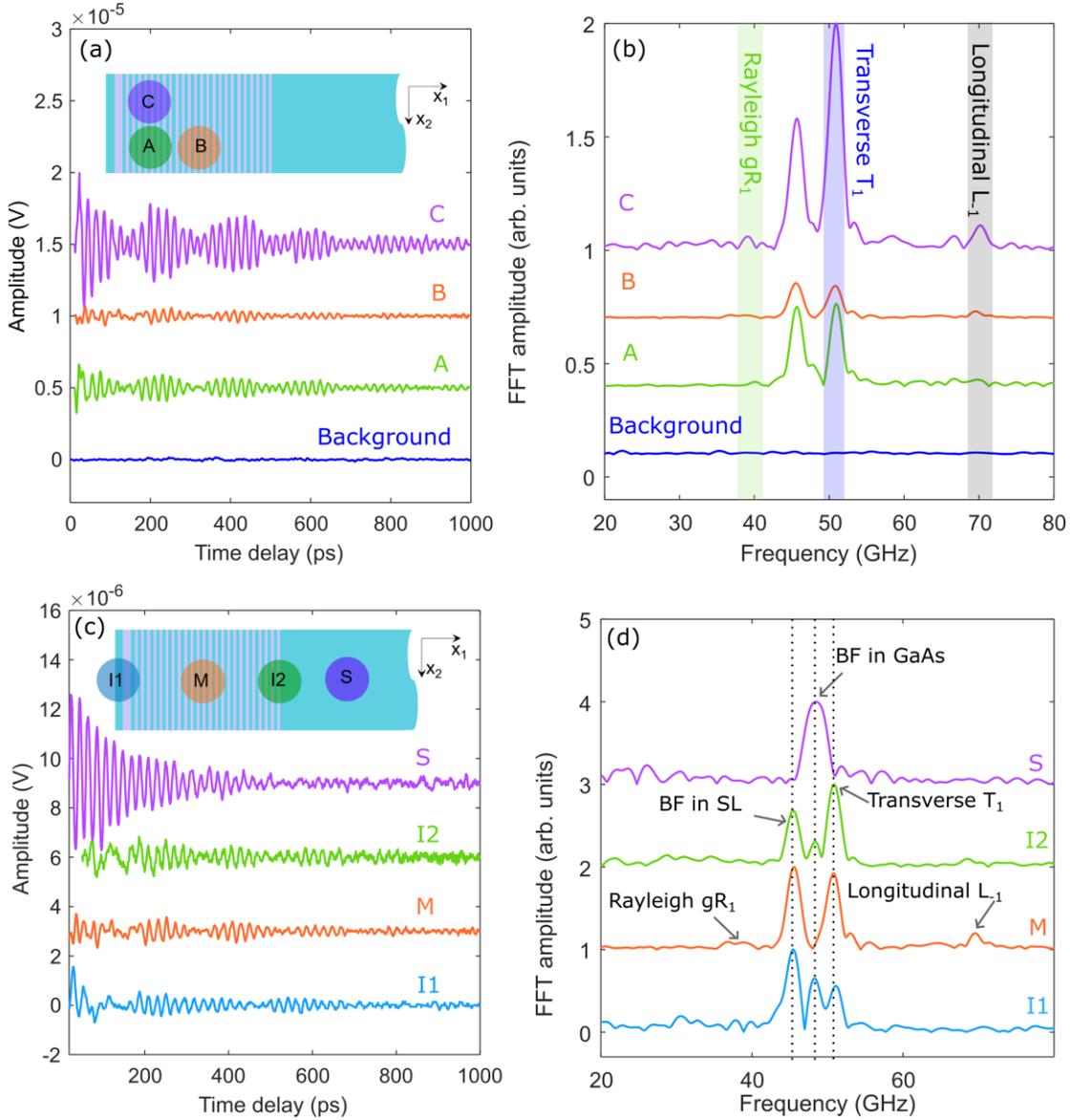

Figure 4. Experimental measurements on the first sample. (a) Detection of acoustic vibrations and (b) their spectra obtained by overlapping the pump and probe foci at positions A, B, C on the SL. These results are compared with background measurements (zero pump light). In (b) the spectral components attributed to the gR, T, and L waves are highlighted in color. (c) Detection of acoustic vibrations and (d) their spectra achieved by overlapping pump/probe foci at positions I1, M, I2, S on top layer/SL interface, SL, SL/GaAs interface, and substrate, respectively. The three spectral peaks indicated by vertical dotted lines correspond to the Brillouin frequency (BF) in the SL (45 GHz, left), the BF in GaAs (48 GHz, middle), and the frequency of the T wave in the SL (51 GHz, right), respectively.

The two additional peaks present in the experimental spectrum in Figure 2c at frequencies of 45.7 GHz and 92.6 GHz can be attributed to the Brillouin frequencies induced by the probe light that is backward scattered by the coherent bulk acoustic modes launched within the bulk (*i.e.*, depth) of the SL nanostructure by the pump laser radiation. This is detailed in Methods H where we derive the analytical estimates of Brillouin frequencies for light scattering by quasi-longitudinal acoustic (QLA) waves propagating in the diffraction orders $n = 0$ and $n = 1$, respectively.

Several spectral components below 40 GHz, not consistently replicable in the experiments, are likely attributable to random noise originating from the optical system or electronic devices. However, it is not within the scope of this discussion to delve further into those frequencies due to their low repeatability.

To assess the reliability and variations in the monitored signals, we adjusted the laser foci laterally and vertically on the cleaved surface. We conducted these adjustments at positions labeled A, B and C within the SL region (as shown in Figure 4a and Figure 4b). Here, we aim to replicate the vibrations induced by the gR, T, and L waves at their anticipated frequencies. However, it is noteworthy that the amplitudes of these vibrations exhibit variations based on the specific positions due to the inherent inhomogeneities within the SL material, as depicted in Figure 1d. Furthermore, these results are compared with the background measurements, and these comparisons indicate that the noise level, approximately at $10^{-7}$, remains consistently below the level of the identified folded acoustic phonons.

It is also enlightening to shift the lateral position of the laser foci across different regions, encompassing the top layer/SL interface (I1), the SL itself (M), the SL/GaAs interface (I2), and the substrate (S) (refer to Figure 4c and 4d). By demarcating three prominent peaks with vertical dotted lines, we uncover that the middle frequency at 48 GHz is notably absent when probing at position M within the SL. However, this frequency is present at interfaces I1 and I2, and corresponds to the Brillouin frequency detected in the GaAs substrate at position S. This observation affirms our expectations that the Brillouin frequency measured within the SL does not coincide with that in the GaAs substrate. The Brillouin frequency at 45 GHz (represented by the left dotted line) detected within the SL region exhibits a red-shift relative to the Brillouin frequency in the GaAs, primarily due to the SL lower refractive index compared to the bulk GaAs when considering the probe light wavelength. The frequency at 51 GHz (as indicated by the right dotted line) is detected at all positions except the substrate, hinting at its origin as a folded acoustic mode arising from the nanostructure periodicity. Additionally, other folded coherent acoustic modes are partially detected at the interfaces (as seen in the comparison between Figure 4b and Figure 4d).

Finally, we investigated the excitation and detection of surface modes on the second sample. A lateral scan of overlapping pump and probe laser beams was conducted on the sample containing the two $Al_{0.2}Ga_{0.8}As/Al_{0.4}Ga_{0.6}As$ SLs (Figure 5a). It is worth noting that the individual thicknesses of $SL_1$, $SL_2$, and the gap are smaller than the focused beam spot. Consequently, the light is always partially absorbed by GaAs when directed at the SLs.

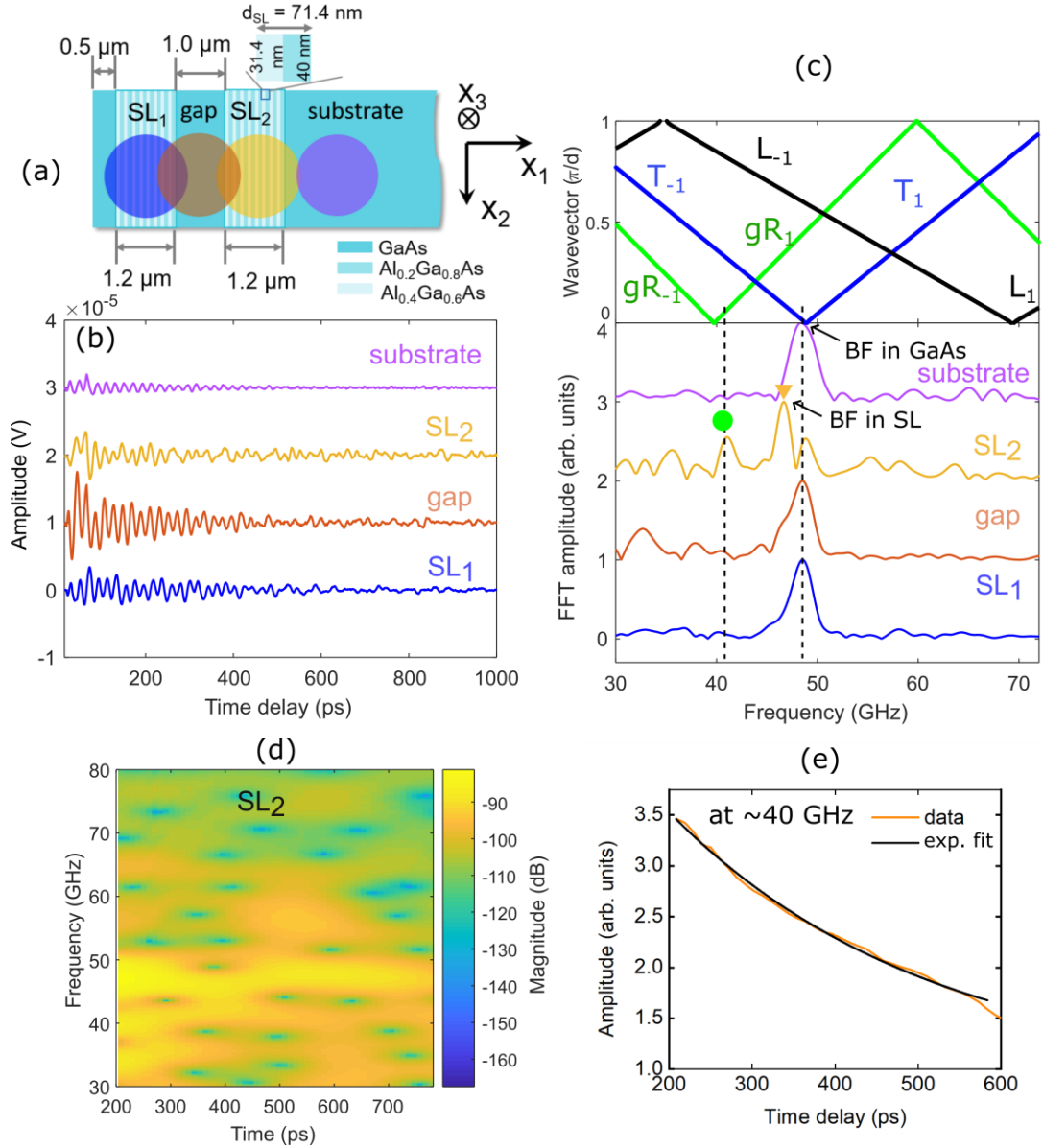

Figure 5. Presentation of the scheme and of the evaluation results for the second sample. (a) Schematic depiction of the sample surface and of the laser foci positions tested on $SL_1$, gap, $SL_2$ and substrate. (b) Contributions to transient reflectivity signal induced by acoustic vibrations. (c) Upper part: computed dispersion relations of acoustic modes in the SLs, considered to be laterally infinite along $x_1$ and semi-infinite along $x_3$ (green: gR, blue: T, black: L); lower part: spectra of acoustic vibrations shown in (b). (d) Spectrogram of the acoustic signal on $SL_2$ shown in part (b) obtained by short-time Fourier transform (see Methods F for details). (e) Attenuation of 40 GHz mode as a function of the time delay, extracted from (d).

The spectral peak at 48.5 GHz is identified as the Brillouin frequency in the GaAs substrate when the laser foci do not overlap with the SL (purple spot in Figure 5a). However, at the three other positions, a potential contribution to this spectral peak from the T wave, theoretically predicted at 49 GHz (see the upper part of Figure 5c) cannot be ruled out.

In SL$_2$, we monitor the folded first-order $k_1 = 0$ gR wave frequency at 41 GHz (indicated by the green solid dot) along with the Brillouin frequency at 46.6 GHz. The absence of these frequency components on the SL$_1$ can be attributed to a non-ideal surface quality or SL quality at the sample edge (there is only 500 nm distance between the top layer/air interface and SL$_1$ left edge).

The peak at about 40 GHz, attributed to the gR mode, exhibits an exponential decay with a lifetime $\tau_S$ of approximately 300 ps, obtained from the exponential fit, shown in Figure 5e, of the amplitude attenuation with time delay at 40 GHz extracted from the spectrogram in Figure 5d. This gR wave lifetime is comparable to that shown in Figure 2d. The propagation velocity of the gR wave can be estimated as $v_{gR} \equiv d_{SL}\omega_{gR}/2\pi \cong 2930$ m/s. The escape time $\tau_a$ of the gR wave from the region where it is monitored, defined by the laser foci radius $a$, can be estimated as $\tau_a \equiv 2a/v_{gR} \approx 550$ ps. Therefore, it is not just the escape time that limits the observed lifetime $\tau_S$ of the gR wave signal. The gR wave lifetime $\tau_{gR}$, potentially limited by gR wave diffraction, its absorption in the bulk of the materials due to inelastic processes, its scattering by surface roughness, and emission of the bulk acoustic waves, can be estimated from the relation $\tau_S^{-1} = \tau_a^{-1} + \tau_{gR}^{-1}$, as $\tau_{gR} \approx 660$ ps. The diffraction time of the gR wave, $\tau_D$, can be estimated as $\tau_D \sim (2a)^2/d_{SL}v_{gR} \approx 12.6$ ns. The time of the longitudinal acoustic wave absorption in the bulk of the GaAs at 56 GHz frequency exceeds 2 ns.[34] In our experiments, the duration of the beatings in acoustically-induced transient reflectivity signals (Figure 2b and Figure 4a,c), resulting from the superposition of signals of comparable amplitudes due to bulk longitudinal acoustic wave and T waves, demonstrates that the lifetimes of the shear acoustic waves at ∼50 GHz are, at least ∼1 ns. Hence, it is expected that neither the diffraction nor the bulk absorption limits the gR wave lifetime in the experiments reported here.

This observation suggests that the gR wave lifetime could be further increased by either reducing the gR wave attenuation caused by surface roughness through improved surface quality (*e.g.*, by depositing an oxide nanolayer to protect the surface from oxidation) or by minimizing gR wave energy emission to bulk modes through the design of an optimized SL structure and composition.

In the second sample, there is no apparent monitoring of L wave, which could be attributed to the reduced contrast in acoustic properties between the constituent layers of Al$_{0.2}$Ga$_{0.8}$As/Al$_{0.4}$Ga$_{0.6}$As SLs in the second sample in comparison to the GaAs/AlAs SL of the first sample. Alternatively, other factors related to sample fabrications, currently not fully understood, might contribute to this observation.

In summary, the transient reflectivity measurements from lateral scanning on the surfaces of both samples can be employed to unveil defects and the nanostructure geometry (including the interfaces of the SL with the surrounding materials) on micrometer scale through the evaluation of acoustic surface vibrations and bulk acoustic eigenmodes. Simultaneously, the monitoring of the gR wave on both samples strengthens the expectations that, in the near future, dedicated studies and optimized designs of the SL-based SAW nanotransducers will enhance the characteristics of the reported gR waves.

It is important to highlight that the signals from monitored T waves are notably prevalent in all our experiments. The following discussion clarifies why this observation has been somewhat unexpected. All forms of bulk acoustic waves, including quasi-longitudinal and quasi-transverse, can be detected using laser-induced gratings in transparent or weakly absorbing materials.[35,36] The detection of L waves was previously reported in multiple experiments utilizing visible[11,36] and UV laser-induced gratings,[19] as well as deposited metallic gratings.[37,38] Concomitantly, these gratings modulate (structure) the near-surface properties of the sample and modulate (ripple) the surface profile. However, there are relatively few reports[38] on T waves detection through laser-induced gratings in opaque materials. Additionally, we are not aware of any demonstrations of T wave generation and detection assisted by deposited metallic gratings.

A potential explanation for the absence of T waves in experiments with elastically isotropic gratings on an elastically isotropic substrate was previously provided.[39] In such elastically isotropic media, the most common laser-induced stresses, *i.e.*, thermoelastic stresses, typically do not generate shear (transverse) acoustic waves within the bulk of the sample.[38] Theoretically, only the transverse mode that is mode-converted from the longitudinal mode, either incident at the surface or propagating along the surface, could contribute to shear surface displacement. Based on accumulated experimental[37,40] and theoretical[39,41] knowledge, it is anticipated that the structures in which either the substrate, the grating, or both exhibit elastic anisotropy, are better suited for controlling T waves. In the absence of elastic anisotropy, gR and L waves are expected to significantly dominate the acoustic signals in transient reflectivity measurements. This rationale aligns with our experimental demonstrations on cleaved SL structures composed of elastically anisotropic constituents. In our experiments, we have successfully accessed both gR and T waves, which feature extremely short but equal spatial periodicity. This characteristic results in a significant separation of their frequencies, even when their propagation velocities are closely matched. This unique platform provides promising opportunities to generate and detect T waves, thus avoiding potential signal interference from the presence of the gR wave. For instance, in isotropic media, measuring the velocity of T waves provides information on the shear rigidity of the material, a property entirely absent in L waves and mixed with information on the longitudinal modulus in the gR wave.[14,42] Unlike L waves, T waves preferentially couple with rotational eigenmodes in materials with complex elasticity.[43,44] In a non-exhaustive list of other advantageous applications of T waves, one can find the determination of residual stress,[45] inspection of material cracks,[46,47] and evaluation of the dynamics of structural phase transitions.[36]

**CONCLUSIONS**

We have successfully showcased the all-optical monitoring of the acoustic vibrations on the samples with periodically structured surface, prepared by cleavage of the SLs along their growth direction. In these structures the periodicity of the material parameters at cleaved mechanically free surface is due to their periodicity in the bulk. These samples, composed of nanometer pitch normally-cleaved $Al_xGa_{1-x}As/Al_yGa_{1-y}As$ SLs, were examined in pump-probe experiments utilizing visible (blue) pump and near-infrared probe laser beams from an ultrafast laser. These semiconductor SLs featured different $Al_xGa_{1-x}As$ constituent layers and a period of 71 nm. We observed and studied folded acoustic phonons, including gR, T, and L waves, within the frequency range of 40 GHz to 70 GHz.

In this tested platform for studying surface acoustic motion at GHz frequencies, T waves dominate over gR waves and L waves. This dominance is tentatively attributed to the elastic anisotropy of the SL. It is worth noting that this may be related to the specific orientation of the SL layering relative to the nanostructure surface, which appears to favor the excitation of T waves by the gradients of the photo-induced stresses. However, it is essential to emphasize that this hypothesis would require further theoretical development, which is beyond the scope of our current presentation, primarily focused on experimental results.

Our platform provides opportunity to monitor T waves and gR waves independently due to the well-separated frequencies by tens of GHz in the nanotransducer. Our proof-of-concept experiments revealed the importance of phenomena such as the oxidation of the freshly cleaved surface, the surface roughness and the non-negligible smallness of the elastic contrast between the layers on the efficiency of SAWs monitoring by laser on the suggested platform. By optimizing these factors along with diminishing the SL period, our platform holds practical opportunities to break the current limitation of SAW frequency, extending it to sub-THz range.

Our experimental findings provide a viable solution for monitoring surface acoustic vibrations with GHz frequencies. This approach serves as a valuable alternative and supplement to established techniques involving metal grating and EUV light grating methods. Unfortunately, our initial trial experiments on the same nanostructured samples, aimed at demonstrating the propagation of the gR wave packets between two spatially separated points on the sample surface by focusing the pump and probe laser beams on these points, expectedly failed. The limited length of gR wave propagation on our SLs, estimated to be approximately $l_{gR} \approx$ 1.9 μm based on the estimated gR wave lifetime and velocity, is comparable to the diameter of the laser beams ($\approx$1.64 μm). These laser beams dictate the lengths of the monitored gR wave packet. Consequently, the pump and probe laser beams essentially overlap, even when the centers of the pump and probe laser beams are separated by the gR wave propagation distance. Moreover, the gR wave propagation time between the focal points is similar to its duration. In simpler terms, the signal increase resulting from the forward movement of the gR wave packet, generated in the pump focus region, toward the detection area controlled by the spatially shifted probe focus, is significantly suppressed or saturated due to the attenuation of gR wave packet happening on the same time scale. This led to gR wave signals falling below the noise level.

Several potential approaches exist for future monitoring of gR wave packets. Firstly, it is desirable to increase the propagation lengths of gR waves. This can be accomplished by reducing surface roughness and preventing surface oxidation. To mitigate oxidation, a protective oxide nanolayer can be applied to the surface immediately after cleaving the SL, or the SLs can be grown from oxide material components. Secondly, shortening the length/duration of the monitored gR wave packets is desirable. This can be achieved by employing stronger laser focusing or by reducing the thickness of the SL to be smaller than the diameter of the laser foci.

Furthermore, it is crucial to combine these approaches with the optimal selection of materials for the SL layers. Material optimization entails enhancing optoacoustic and acousto-optic transformation efficiencies at specific wavelengths of the pump and probe lasers. By improving these efficiencies, the amplitude of the signal resulting from coherent gR waves would increase, potentially elevating gR wave amplitudes above the noise level at micrometer propagation distances.

Sub-THz SAWs are potentially promising for future applications in science and technology, such as the next generation of ICT devices well beyond the 5G standard (600 MHz - 6 GHz). The recent demonstration of picosecond acoustics based on miniaturized lasers[16] represents a significant step forward in applying laser-monitored SAWs to future ICT. A detailed understanding of the propagation of such SAWs on atomically flat surfaces and across material junctions is a prerequisite for accessing the fundamental concepts of thermal load management. With the diminishing periodicity of the SL, the platform of the cleaved SLs presented herein also holds considerable potential for revealing and exploring strong acousto-optic and opto-acoustic interactions due to the simultaneous confinement of acoustic and electromagnetic modes.

**METHODS**

**A. Sample Preparation and Characterization by X-ray Diffraction and AFM**

The SL samples were grown using solid-source molecular beam epitaxy (MBE) with a Veeco GenXplor system. The $Al_xGa_{1-x}As$ alloy was achieved through a digital alloy growth technique, where the composition is determined by the layer thicknesses of GaAs and AlAs. The samples were grown under typical conditions for GaAs and AlAs, maintaining an As-rich environment at 580 °C. The growth rates were approximately 1 ML/sec for both GaAs and AlAs. High-resolution XRD scans on the (004) plane were conducted using a Bruker D8 system. AFM measurements were performed with a 5500 AFM from Agilent® in intermittent contact mode in air (cantilever stiffness was 40 N/m). Image processing utilized Gwyddion® freeware.

**B. Structure of the Generalized Rayleigh SAWs in Cleaved SLs**

For the physical interpretation of the demonstrated all-optical monitoring of the Rayleigh-type SAW (R) wave on the SL stratified normal to the surface, it is important to emphasize the key differences in the mode structure of this gR wave compared to the Rayleigh SAW (RSAW) on the surface of a homogeneous half-space. As explained by the Bloch-Floquet theorem,[48] there is an infinite number of the bulk longitudinal and transverse wave eigenmodes in the SL presented in Figure 2c. This is attributed to the periodicity of SL parameters, characterized by the spatial period $d_{SL}$ or by the SL (Floquet) wave number $q_{SL} = 2\pi/d_{SL}$ along the $x_1$ axis (the direction of SL layering).

The $m$-th order bulk eigenmodes can be viewed as the diffraction orders of the bulk modes with frequency $\omega$ and the wave vector $\vec{k}$ (with $k_2 = 0$ for the waves polarized in the sagittal plane $(x_1 x_3)$), which differ by the projections of $k_1$ ($k_1 + mq_{SL}, m = 0, \pm 1, \pm 2, ...$) on the $x_1$-axis. Crucially, the displacement components with different $m$ are parametrically coupled due to the periodicity of the acoustical properties in the SL. Consequently, each eigenmodes, numbered as the $m$-th mode, contains components with all allowed $m = 0, \pm1, \pm2, ...$, including, in general, $m = 0$, although with different amplitudes.

The gR wave results from the superposition of the bulk modes, evanescent in the depth direction $x_3$ (*i.e.*, with $k_3$ describing amplitude decay with depth) and satisfying periodic stress-free boundary condition on $x_3 = 0$ at a particular gR wave eigenfrequency, $\omega_{gR}$. It contains displacement components with all $k_1 + mq_{SL}, m = 0, \pm1, \pm2, ...$ (see Supporting Information S1 for details). Similar to the dispersion relations for bulk acoustic waves, the gR wave dispersion relation can be folded in the Brillouin zone with the acoustic phonon branches, starting from $\omega = 0$, and the optical phonons branches[28,49]. The important difference of the gR wave from RSAW is that the former contains, in the zone-center $k_1 = 0$, laterally non-modulated displacement and strain components at the frequencies of optical-type SAWs.

### C. Photogeneration of the Coherent Acoustic Eigenmodes in Cleaved SLs

In our experiments, the phonons in the close vicinity of the Brillouin zone center, *i.e.*, with $k_1 \ll q_{SL} \equiv 2\pi/d_{SL}$, play the dominant role, as evident from the fact that the radii of pump and probe laser beams foci on the SL surface, $a_{pump} \approx a_{probe} \approx a$, significantly exceed the SL period $d_{SL}$. The $\vec{k}$-spectrum of the photo-generated SAWs is controlled by the lateral distribution of the stresses, photo-induced by the pump laser pulses (see Supporting Information S1 for the details). In our experiments, the lateral modulation of the photo-induced stresses on the nanometer spatial scale is mostly due to the difference in the optical properties of the two layers.

In the infinite SL, the stresses photo-induced by laterally homogeneous pump laser radiation will have the periodicity of the SL and exhibit the Bloch-Floquet property. Thus, the folded $k_1$-spectrum of the photo-induced stresses would contain only zone center components, $k_1 = 0$. Focusing the pump laser radiation in a spot with radius $a$ localizes laterally the photo-induced stresses on the scale $a \gg d_{SL}$, even in the case of a diffraction-limited focusing, because of the sub-optical periodicity of the SL, *i.e.*, $d_{SL}/\lambda \ll 1$. Therefore, the related broadening of the $k_1$-spectrum components is much narrower than the width of the Brillouin zone $\pi/d_{SL}$ and the photo-generated gR wave spectrum is strongly localized near the Brillouin zone center.

The condition $a \gg d_{SL}$ also leads to the expectations that symmetric motions of individual SL layers are photo-excited with larger amplitudes than antisymmetric ones. Here, we denote symmetric modes as modes that exhibits an odd distribution of the horizontal mechanical displacement, $u_1$, and an even distribution of vertical mechanical displacement, $u_3$, relative to the central plane of the individual layer (see Supporting Information S1 for the details).

For the experimental platform of a semi-infinite SL stratified normal to the surface, it is crucial that the zone-center surface confined modes contain laterally homogeneous components in their structure. This implies that gR modes can be excited by a laterally homogeneous (averaged over the SL period) part of the distributed photo-induced stresses or, in principle, even by laterally non-modulated stresses in the hypothetical case of the absence of optical but existing acoustical contrast between the SL composing layers.

Under the condition of weak acoustical contrast, $\Delta\rho/\langle\rho\rangle \sim \Delta C_{ij}/\langle C_{ij}\rangle \sim \mu \ll 1$, which is sufficiently well satisfied in our experimental samples, the amplitude ratio of the non-modulated to modulated zone-center gR wave components scales with $\sim\mu$, the frequency of the lowest zone center optical-type gR wave, $\omega_{GR}$, deviates from the respective frequency $\omega_R$ of RSAW with the wave vector $k_1 = 2\pi/d_{SL}$ in the medium of the period-averaged parameters, $\text{Re}(\omega_{gR} - \omega_{gR})/\omega_{gR} \sim \mu^2$, and it acquires an imaginary part $\text{Im}(\omega_{gR})/\omega_{gR} \sim \mu^2$. Thus, the zone center gR waves are actually pseudo-SAWs[50,51].

### D. Photodetection of the Coherent Acoustic Eigenmodes in Cleaved SLs

The laterally homogeneous components of the high frequency acoustic eigenmodes in cleaved SLs can be detected by probe laser radiation, as established in picosecond laser ultrasonics of depth-stratified samples/structures.[52] The theoretical analysis of the detection paths of gR wave, when it is propagating in the SL (identical to the one where it was photo-generated), reveals that the laterally modulated components of the gR wave can also be efficiently detected due to the lateral periodicity in the SL of the optical permittivity $\varepsilon$ and/or the photoelastic parameter $p$ (see Supporting Information S2 for the details).

The essence of the acoustic wave detection in the ultrafast pump-probe optical experiments (via measurements of the transient reflectivity at normal incidence of probe laser light) is the heterodyning of the weak acoustically-scattered probe light by the strong probe light reflected from the surface of the SL[53]. Only the probe light reflected in the $m = 0$ diffraction order reaches the photodetector because the light reflected in the other diffraction orders is evanescent due to sub-optical periodicity of the considered cleaved SLs. Consequently, only the probe light scattered by the acoustic wave in the zeroth diffraction order is detectable.

An additional consequence of the sub-optical periodicity of the SL is that each diffraction of light by the modulated components of the SL parameters acquires a multiplier $(d_{SL}/\lambda)^2 \ll 1$ in the amplitude of the diffracted light. Because of this, the non-evanescent probe light that can propagate in the direction of the photodetector (in the zeroth diffraction order) after exhibiting several consecutive diffractions (in reflection and/or scattering) is negligibly small in amplitude compared to the non-diffracted probe light.

Only the scattering of the non-diffracted probe light by the acoustically-induced modulation of the period-averaged optical permittivity $\langle\varepsilon\rangle$ of the SL should be considered. However, this laterally homogeneous modulation is caused not only by the modulation of the averaged permittivity by the laterally non-modulated gR wave component (first path) but also by the demodulation of the laterally modulated components of the gR wave when it interacts with SL parameters, exhibiting the same periodicity as the gR wave (second path). From the viewpoint of formal mathematics, the second path is due to the presence of the constant/infinite-period term in the square of the sine/cosine functions. In particular, the second path of the possible detection can be due to $\varepsilon^{(1)} - \varepsilon^{(2)} \neq 0$ and/or $(\varepsilon^{(1)})^2 p^{(1)} - (\varepsilon^{(2)})^2 p^{(2)} \neq 0$, *i.e.*, due to optical or/and acousto-optical contrast between the SL layers (see Supporting Information S2).

Note that, in the case $\varepsilon^{(1)} - \varepsilon^{(2)} \neq 0$, the second detection path can be due to the difference in the individual layer thickness variations induced by horizontal displacements components of the gR wave even in the case of the negligible photoelasticity of both layers ($p = 0$). The theory indicates that either the first or the second detection path could dominate depending on the relative strengths of the acoustical (density $\rho$, elastic moduli $C_{ij}$,) or optical/acousto-optical (permittivity, photoelastic moduli) contrasts between the SL layers. The first path progressively disappears with diminishing acoustic contrast because of the disappearance of the laterally unmodulated component of gR wave. However, it could still be the dominant path in the hypothetical case of the absence of optical and acousto-optical contrasts (see Supporting Information S2 for the details).

Overall, the inevitable diminishing of the detected gR wave signals with the diminishing of the sub-optical SL pitch ($\sim d_{SL}/\lambda$, see Supporting Information S2) is not the direct consequence of the fall in light diffraction efficiency due to nanometer scale periodicity but is caused by the diminishing of the acousto-optic scattering volume proportionally to the penetration depth of the gR wave, $\sim d_{SL}$.

### E. Ultrafast Pump-Probe Laser Experiments

The acoustic vibrations are monitored through ultrafast pump-probe experiments employing a femtosecond laser. A Ti: sapphire laser (Spectra-Physics) with a central wavelength of 810 nm, a pulse width of ~100 fs, and a pulse repetition rate of ~80 MHz is utilized. The laser beam at the fundamental wavelength probes the sample, while the laser beam at the second harmonic, generated in a BBO crystal, is employed to pump/excite the sample. A translation stage inserted in the pump path induces the pump-probe delay, with a maximum range of ~8 ns.

The attenuated pump and probe laser beams are collinearly focused by an objective (100×, NA of 0.55) on the same surface of the sample. The reflected probe light is detected by a large-area 150 kHz-bandwidth photoreceiver (NewFocus, model 2031), whose output is connected to a lock-in amplifier for synchronous detection at the 150 kHz modulation frequency of the pump intensity. This allows for low-noise signal acquisition, transmitted to an analog-to-digital convertor for data collection. To test different lateral positions on the sample surface, the sample is displaced relative to laser foci by a translation stage.

### F. Signal Processing Methods

*(a) Background deletion.* The time-dependent reflectivity change undergoes numerical processing to reveal small-amplitude acoustic oscillations. These oscillations are extracted by subtracting the non-oscillating background from the raw signal. The background is obtained using a moving average method with a Gaussian window duration of approximately 100 ps. Removing the background does not introduce any new frequency appearances or alterations; instead, it supresses undesired relatively low frequency components or environmental/mechanical/thermal noises.

*(b) Fourier transform.* The spectra are obtained by applying fast Fourier transform (FFT) to the time-domain acoustic signals. A Hanning window is multiplied over the signal of interest for FFT.

*(c) Short-time FFT.* To observe time-dependent changes in spectra amplitude (Figure 2d) and for the second sample (Figure 5d), a short-time FFT procedure is used. During the procedure, the parameters include Gaussian window of 400 ps duration, a frequency resolution for frequency axis of 0.1 GHz, and a hop time of 10 ps.

### G. Finite Element Model (dispersion relation and acoustic eigenmodes)

The dispersion curves of surface acoustic waves (SAWs) propagating at the free surface of assumed infinitely cleaved superlattices are computed using finite element method (FEM) with eigenfrequency analysis in COMSOL Multiphysics.

*(a) Structure, material and boundaries*. A unit of the superlattice, consisting of alternating two layers with a period of $d_{SL}$ is constructed (see Supporting Information S1 for details on the structure). The height of the superlattice along the $x_3$ direction is set to $20d_{SL}$. The [100] crystallographic orientation of constitutive materials is aligned with the $x_1$ axis (SL growth direction) in the calculation. Since the displacement component normal to the $x_1x_3$-plane equals zero, the model is treated as a 2D problem. The two layers are considered as linear elastic cubic materials based on $Al_xGa_{1-x}As$. Floquet periodic boundary conditions are applied to the left and right sides of the unit cell along the $x_1$ direction. The top surface is assumed to be a stress-free boundary. A perfectly matched layer (PML) is placed at the bottom of the unit cell to simulate an infinite half-space. The bottom surface of the unit cell is set as a fixed boundary to limit the acoustic eigenmodes localized at that bottom surface. The PML length is set as one period of the SL.

*(b) Calculation*. The solid mechanics interface of the structural mechanics module of the COMSOL Multiphysics is used to solve an eigenfrequency problem based on the elastic Helmholtz equation. The eigenfrequencies are searched through a parametric sweep over the first Brillouin zone $k_1 \in [0, \pi/d_{SL}]$ with a step of $\pi/d_{SL}/100$. The eigenvalue solver is MUMPS (multifrontal massively parallel sparse direct solver), set to look for 1000 eigenfrequencies to find all the relevant eigenvalues. The eigen vectors for each solution correspond to the mode shape, including the horizontal displacement $u_1$ (along $x_1$-axis) and the vertical displacement $u_3$ (along $x_3$-axis). The computation time for these settings was 6h33min on a computer (Intel core i9-9980HK CPU@2.4 GHz).

*(c) Mode sorting*. First, solutions with a large imaginary part of its eigenfrequency are excluded. Second, the mode shape (displacement field or energy spatial distribution) of the remaining solutions is used to distinguish the gR wave (confined close to the surface, see Figure 3a), T wave (with dominant vertical displacement, see Figure 3b) and L wave (with dominant horizontal displacement, see Figure 3b) modes. At last, for validation, the obtained mode branches are compared to their expected phase velocities.

### H. Brillouin Light Scattering in Cleaved SLs

The efficient backward scattering of the probe light in the zeroth order of diffraction, the only non-evanescent light field component in the deeply sub-optical SL, is governed by the momentum conservation conditions $k_1 = q_1 = 0$ and $k_3 = 2q_3$, where $k_i$ and $q_i$ denote the components of the acoustic and probe light wave vectors, respectively. The normal components $k_3$ of the coherent acoustic waves in the different diffraction orders $m$ are given by $k_{3,QTA,QLA} = \sqrt{k_{QTA,QLA}^2 - (2\pi m/d_{SL})^2}$, where $k_{QTA,QLA} = \omega/v_{QTA,QLA}$ are the wave vectors of the quasi-transversal (QTA) and quasi-longitudinal (QLA) acoustic waves propagating along the directions of the respective diffraction orders in the SL, $d_{SL}$ is the SL period and $m = 0, \pm 1, \pm 2, \ldots$ Therefore the velocities $v_{QTA,QLA}$ of the QTA and QLA waves in the above formula should be evaluated along the directions defined by the following angles $\gamma \equiv \gamma(d_{SL}, n) = \tan^{-1}[(2\pi m/d_{SL})/k_{3,QTA,QLA}] = \tan^{-1}[(2\pi m/d_{SL})/(2q_3)]$ relative to the $x_3$ axis. Thus, the knowledge of the probe light wave vector and the slowness curves (anisotropy) of the acoustic modes in the SL provides an opportunity to estimate the Brillouin frequencies.

In our estimates, as described in Supporting Information S1 and based on the weak elastic contrast between the layers, we used the dependences of the acoustic velocities on the propagation direction/angle provided by the slowness curves of the GaAs[42] and averaged optical refractive index $\langle n \rangle$. Two of the spectral peaks in the lower part of Figure 2c can be tentatively attributed to Brillouin scattering of probe by QLA waves in the orders $m = 0$ and $m = 1$, respectively. All these waves are non-evanescent. Note that symmetry considerations prevent the generation of the QTA waves with the order $m = 0$ in our experimental configuration.[42] It is also worth noting that the SL is optically birefringent,[54] and thus, potentially each of the Brillouin peaks could be composed of up to three neighboring spectral components, corresponding to different combinations of ordinary and extraordinary probe light propagating from and towards the SL surface.[55] We have experimentally verified that the position of the most intense Brillouin peak around 45.7 GHz (Figure 2c) depends on the polarization of the probe light. However, the splitting of the peak is not resolved in our experiments because of weak birefringence, which is caused only by structuring, while the materials of the SL layers are optically isotropic.

## ASSOCIATED CONTENT

**Supporting Information**
Structure of generalized Rayleigh-type surface acoustic wave on a superlattice stratified normal to its surface; Optical detection of generalized Rayleigh-type surface acoustic wave on a superlattice stratified normal to its surface.

**Author Contributions**
C.L. conducted the experiments, executed the simulations and input corresponding sections for the manuscript. N.C conducted the first experimental tests and helped with the experiments. T.T. participated the first trial experiments. V.T helped with the initial simulations. N.D conducted the AFM characterization. H.L and K.Z. prepared and characterized the samples by XRD. S.R. assisted with signal processing, improved the manuscript. V.G. and H.L designed the samples. V.G. designed the methodology, wrote the first manuscript and supervised the project. All authors reviewed the manuscript.

**Notes**
The authors declare no competing financial interest.

## ACKNOWLEDGEMENT

This research is supported by the postdoctoral fellowships of the Institut d'Acoustique – Graduate School (IA-GS) of Le Mans Université and of European Commission's Horizon 2020 research and innovation programme under the Marie Skłodowska-Curie grant agreement No. 101025424 for C. Li.## ABBREVIATIONS

RSAW, Rayleigh surface acoustic wave; SL, superlattice; SAW, surface acoustic wave; OA, optoacoustic; AO, acousto-optic; OAOT, opto-acousto-optic transducer; EUV, extreme ultraviolet; NIR, near infrared; MBE, molecular beam epitaxy; XRD, X-ray diffraction; AFM, atomic force microscopy; gR, generalized Rayleigh surface acoustic wave; T, skimming surface transverse acoustic wave; L, skimming surface longitudinal acoustic wave; QLA, quasi-longitudinal acoustic; QTA, quasi-transversal acoustic.

# Optically-Controlled Nano-Transducers Based on Cleaved Superlattices for Monitoring Gigahertz Surface Acoustic Vibrations

Changxiu Li[1], Nikolay Chigarev[1], Théo Thréard[1], Kedong Zhang[2], Nicolas Delorme[3], Vincent Tournat[1], Samuel Raetz[1], Hong Lu[2], Vitalyi E. Gusev[1,*]

[1]Laboratoire d'Acoustique de l'Université du Mans (LAUM), UMR 6613, Institut d'Acoustique – Graduate School (IA-GS), CNRS, Le Mans Université, Le Mans, France
[2]College of Engineering and Applied Sciences, Nanjing University, Nanjing, China
[3]Institut des Molécules et Matériaux du Mans (IMMM), UMR 6283 CNRS, Le Mans Université, Le Mans, France

* vitali.goussev@univ-lemans.fr

CONTENT

S1. Structure of generalized Rayleigh-type surface acoustic wave on a superlattice stratified normal to its surface

S2. Optical detection of generalized Rayleigh-type surface acoustic wave on a superlattice stratified normal to its surface

# Structure of generalized surface acoustic Rayleigh wave on a superlattice stratified normal to its surface

The mathematical formalism for the theoretical investigation of the surface acoustic waves (SAWs) which could propagate in the direction normal to the layers of the superlattice (SL) along mechanically free surface which is also normal to SL layers (Fig. S1) was developed nearly 40 years ago [S1]. In Fig. S1 a bilayer SL with individual infinite plane layers parallel to $x_2$ axis and the layering normal to $x_1$ axis is schematically presented. In general, the layers are of different thicknesses $a^{(1)}$ and $a^{(2)}$ and are of materials with different densities $\rho^{(1,2)}$ and elastic moduli tensors $C_{\alpha\beta\gamma\delta}^{(1,2)}$.

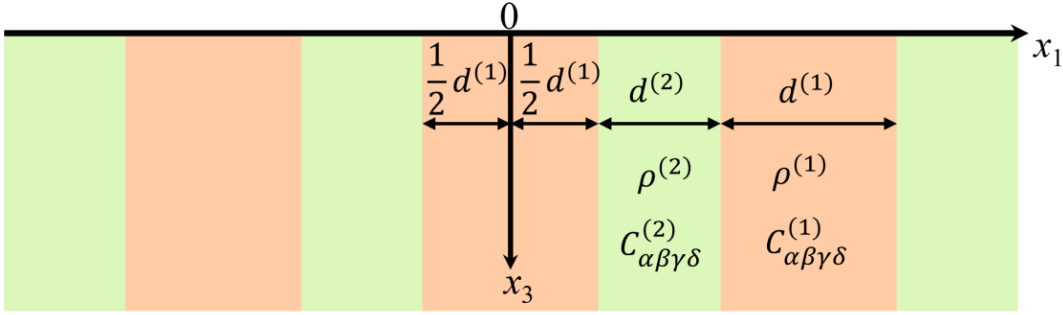

**Fig. S1**. Scheme of a bilayer superlattice stratified normal to the mechanically free surface.

In the geometry presented in Fig. S1 the generalized Rayleigh-type SAWs, which are denoted in the main text of the manuscript as gR waves, are polarized in the sagittal plane ($x_1$,$x_3$). In an acoustic wave of cyclic frequency $\omega$ two particle displacements components $u_\alpha(\vec{x}, t) = u_\alpha(\vec{x})e^{-i\omega t}$, where ($\alpha = 1,3$), satisfy the system of two coupled Helmholtz equations $-\omega^2 \rho(\vec{x}) u_\alpha = \frac{\partial}{\partial x_\beta}\left[C_{\alpha\beta\gamma\delta}(\vec{x})\frac{\partial u_\gamma}{\partial x_\delta}\right]$. Note, that the density and the elastic moduli in these equations are periodic functions of $x_1$ coordinate and are independent of two other coordinates. In [S1] the Helmholtz equations were specified for the case where the materials composing the SL are both of cubic symmetry, with crystallographic axes oriented along the coordinate axes in Fig. S1:

$$-\omega^2 \rho(x_1) u_1(\vec{x}) = \frac{\partial C_{11}(x_1)}{\partial x_1}\frac{\partial u_1}{\partial x_1} + \frac{\partial C_{12}(x_1)}{\partial x_1}\frac{\partial u_3}{\partial x_3} + [C_{12}(x_1) + C_{44}(x_1)]\frac{\partial^2 u_3}{\partial x_1 \partial x_3} + C_{11}(x_1)\frac{\partial^2 u_1}{\partial x_1^2} + C_{44}(x_1)\frac{\partial^2 u_1}{\partial x_3^2},$$

$$-\omega^2 \rho(x_1) u_3(\vec{x}) = \frac{\partial C_{44}(x_1)}{\partial x_1}\frac{\partial u_1}{\partial x_3} + \frac{\partial C_{44}(x_1)}{\partial x_1}\frac{\partial u_3}{\partial x_1} + [C_{12}(x_1) + C_{44}(x_1)]\frac{\partial^2 u_1}{\partial x_1 \partial x_3} + C_{11}(x_1)\frac{\partial^2 u_3}{\partial x_3^2} + C_{44}(x_1)\frac{\partial^2 u_3}{\partial x_1^2}.$$

The boundary conditions at stress-free surface $x_3 = 0$ are:

$$\left[C_{44}(x_1)\left(\frac{\partial u_1}{\partial x_3} + \frac{\partial u_3}{\partial x_1}\right)\right]_{x_3=0} = 0, \qquad \left[C_{12}(x_1)\frac{\partial u_1}{\partial x_1} + C_{11}(x_1)\frac{\partial u_3}{\partial x_3}\right]_{x_3=0} = 0.$$

These equations and boundary conditions are valid for all the SLs in our experiments. In [S1] the mathematical formalism for revealing the Rayleigh-type SAWs was developed for an arbitrary periodic distributions of density $\rho(x_1)$ and elastic moduli $C_{11}(x_1), C_{12}(x_1)$ and $C_{44}(x_1)$, expanding them in the Fourier series

$$f(x_1) = \sum_{m=-\infty}^{\infty} f(m) e^{imq_{SL}x_1}, \quad f(m) = \frac{1}{a} \int_{-d_{SL}/2}^{d_{SL}/2} dx_1 f(x_1) e^{-imq_{SL}x_1}, \tag{S1}$$

Where $f$ denotes either the density or the elastic modulus, $d_{SL} \equiv d^{(1)} + d^{(2)}$ and $q \equiv 2\pi/d_{SL}$ denote the SL period and SL wave number, respectively. In accordance with Bloch-Floquet property of the waves in the spatially periodic media [S2] the solution of the Helmholtz equations for the bulk eigen modes has the following form:

$$u_\alpha(x_1, x_3) = \sum_{m=-\infty}^{\infty} u_\alpha(m, x_3) e^{ik_m x_1}, \tag{S2}$$

where $k_m = k + q_{SL} m$. Thus, each component of the mechanical displacement is a periodic function along $x_1$ axis composed of the component with wave number $k$, corresponding to $m = 0$ in Eq. (S2) and independent of the SL periodicity, and an infinite number of the components with wave numbers shifted relative to $k$ by a $q_{SL} m$ ($m = \pm 1, \pm 2, ...$). Saying differently, each component $u_\alpha(x_1, x_3)$ of the mechanical displacement vector is composed of an infinite number of contributions from all the « diffraction orders » $m$. The periodic inhomogeneity of the material couples all the « diffraction orders » and it couples different components of the mechanical displacement additionally to their coupling by the boundary conditions. Substitution of the above-assumed spatially periodic distributions of the material parameters, Eq.(S1), and of the mechanical displacements, Eq. (2), in the system of the Helmholtz equations results in the infinite system of coupled algebraic equations for $u_\alpha(m, x_3)$ [S1]. For the particular frequencies $\omega$, the evanescent bulk eigenmodes, i.e., those decaying with increasing distance $x_3$ from the surface ($x_1, x_2$), can be combined with particular relative amplitudes to satisfy the conditions at stress-free surface. In [S1] an infinite system of equations for the amplitudes of the bulk components contributing to the surface acoustic wave was derived. These equations are compatible only for the particular values of frequencies, which are the eigen frequencies of the SAWs. Because all the wave numbers $k_m = k + qm$ of the displacement field are equivalent, the dispersion relations $\omega = \omega(k)$ of the acoustic waves in the structure presented in Fig. S1 can be folded inside the Brillouin zone $\pi/d_{SL} \leq k \leq \pi/d_{SL}$, where $\pi/d_{SL} \equiv k_{edge}$ denotes the Brillouin zone edge (Fig. S2 and Fig. 2 (c) of the manuscript).

The derived systems of algebraic equations were solved in [S1] only numerically and only non-attenuated SAWs, i.e., with real valued frequencies, were in the scope of [S1]. Thus, mostly the lowest (acoustical-type) branch of the SAW dispersion relation was evaluated from the zone center, where for this mode $\omega(k = 0) = 0$, to the zone edge $k = k_{edge}$. The possibility of the gap opening between acoustical-type SAW branch and the nearest in frequencies (first) optical-type SAW branch at the Brillouin zone edge was revealed. The first optical-type SAW branch was studied in [S1] only rather close to the zone edge. With the diminishing wave number, the first optical-type SAW branch intersects the lowest (acoustical-type) branches of the bulk modes, e.g. Fig. S2 and Fig. 2(c) of the manuscript. At wave numbers smaller than this intersection point, the SAW is supersonic relative to some of the bulk modes and emits energy in the bulk of the sample. Purely real solution for the SAW eigen frequency disappears. The attenuated SAWs, including optical-type SAW modes in the zone center ($k = 0$) and nearby, which are playing the dominant role in our experiments, were not studied in [S1] at all.

In our experiments, the $k$ spectrum of the photo-generated SAWs is controlled by the spatial distribution of the stresses that could be photo-induced by the pump laser pulses. Of course, in general, the distribution of the photo-induced stresses depends on the physical mechanisms of optoacoustic conversion (thermoelasticity, electron-phonon deformation potential, inverse piezoelectric effect, electrostriction, etc. [S3, S4]). The dominant physical mechanisms in each particular case are determined by the physical properties of the materials composing the SL and the energy quanta of the pump laser radiation. For example, the pump light penetration depth in the materials composing the SL can importantly influence the depth distribution of the photo-induced stresses [S3, S4]. The universal influence on the lateral distribution of the photo-induced stresses, i.e., along the free surface, and, respectively, on the $k$ spectrum of the photo-induced stresses, stems from the lateral distribution of the pump laser radiation penetrating inside the SL. In our experiments, the lateral modulation of the photo-induced stresses at nanometer spatial scale is due to the difference in the optical properties of the two layers. In the infinite SL composed of spatially homogeneous layers the stresses photo-induced by the laterally homogeneous incident pump laser radiation will have the periodicity of SL and will exhibit the Bloch-Floquet property. Thus, in this limiting case the folded $k$ spectrum of the photo-induced stresses would contain only zone center components, $k = 0$. Focusing of the pump laser radiation localizes initial optical excitation laterally at the scale which is larger or comparable to the pump optical wavelength. Consequently, in our experiments the $k$ spectrum of the photo-induced stresses would be broadened. However, in our experiments, the period of the SL, $d_{SL}$, is much smaller than the optical wavelengths. Therefore, this broadening is much narrower than the width of the Brillouin zone $\pi/d_{SL} \equiv k_{edge}$. Consequently, the photo-generated SAWs $k$ spectrum is expected to be strongly localized near the Brillouin zone center $k = 0$.

In order to get a physical insight relevant to our experiments, we drastically simplified the general approach for the description of the gR waves [S1]. Firstly, based on the above-described arguments, we evaluate analytically only $k = 0$ modes. Secondly, assuming the layers of equal thickness, we take into account only the first two lowest order components of the waves, i.e., corresponding to $m = 0$ and $m = \pm 1$, in Eq. (S2). This simplification is based on the following arguments. If $d^{(1)} = d^{(2)}$, then only $m = 0$ and odd $m$ contribute to the Fourier spectra of the material parameter distributions in Eq. (S1), while the Fourier components with $|m| > 1$ are significantly smaller than $|m| = 1$ components. The same is valid for the lateral distribution components of the pump laser intensity penetrating inside the infinite SL because they are controlled by the periodic distribution of optical reflection coefficient. Consequently, the dominant components in the SAWs are expected to be $m = 0$ and $|m| = 1$, coupled inside the SL by the Fourier components $|m| = 1$ of the density and elastic moduli distributions. Thirdly, as the laser spot size on focal plane is much larger than the SL period, the lateral variations of the photo-induced stresses across the individual layer thickness are small in comparison with the stresses themselves. This leads to the expectations that the symmetric motions of the individual SL layers are photo-excited with larger amplitudes than antisymmetric ones. A symmetric mode exhibits, relative to the central plane of the individual layer, an odd distribution of the horizontal mechanical displacement, $u_1$, and an even distribution of vertical mechanical displacement, $u_3$. Combining the above assumptions, we search the solutions of the Helmholtz equations for the bulk modes in the following simplified form:

$$u_1(x_1, x_3) = 2u_S \sin(q_{SL} x_1) e^{-\alpha x_3}, u_3(x_1, x_3) = [u_0 + 2u_C \cos(q_{SL} x_1)] e^{-\alpha x_3},$$

$$f(x_1) = \langle f \rangle + 2\Delta f \cos(qx_1), \langle f \rangle = (f^{(1)} + f^{(2)})/2, \Delta f = (f^{(1)} - f^{(2)})/\pi. \tag{S3}$$

Here $\alpha$ denotes the depth propagation/penetration constant [S1], related to the projection of the acoustic wave vector on the $x_3$ direction. Substitution of the assumed displacement field Eq. (S3) and the spatial distributions of density $\rho(x_1)$ and elastic moduli $C_{11}(x_1), C_{12}(x_1)$ and $C_{44}(x_1)$ in the Helmholtz equations and separation of the constant terms and the terms proportional to $\sin(q_{SL} x_1)$ and to $\cos(q_{SL} x_1)$, results in the system of equations for the displacement amplitudes $u_S, u_0$ and $u_C$:

$$\begin{pmatrix} [\omega^2 \rho + \alpha^2 C_{44} - q_{SL}^2 C_{11}] & \alpha q_{SL}(C_{12} + C_{44}) & \alpha q_{SL} \Delta C_{12} \\ -\alpha q_{SL}(C_{12} + C_{44}) & [\omega^2 \rho + \alpha^2 C_{11} - q_{SL}^2 C_{44}] & [\omega^2 \Delta\rho + \alpha^2 \Delta C_{11}] \\ -\alpha q_{SL} \Delta C_{12} & [\omega^2 \Delta\rho + \alpha^2 \Delta C_{11}] & \frac{1}{2}[\omega^2 \rho + \alpha^2 C_{11}] \end{pmatrix} \begin{pmatrix} u_S \\ u_C \\ u_0 \end{pmatrix} = \begin{pmatrix} 0 \\ 0 \\ 0 \end{pmatrix}.$$

The compatibility condition for these equations defines the bulk eigen modes in the geometry presented in Fig. S1. Although the compatibility condition can be solved exactly, we use here the method of the successive approximations, based on the following assumption of a weak modulation of the material parameters: $\Delta\rho/\langle\rho\rangle \sim \Delta C_{12}/\langle C_{12}\rangle \sim \Delta C_{11}/\langle C_{11}\rangle \sim \mu \ll 1$. This inequality is satisfied sufficiently well in our samples and leads to very insightful estimates. Additionally, to avoid rather cumbersome formulas and get even more insightful predictions, we assume that the SL is composed of elastically isotropic layers, $C_{12} = C_{11} - 2C_{44}$. Therefore, relations between the elastic moduli and the velocities of the bulk acoustic waves are significantly more compact than in the cubic crystals: $C_{11} = \rho C_L^2$ and $C_{44} = \rho C_T^2$. Here $C_L$ and $C_T$ denote the velocities of the longitudinal and the transverse bulk acoustic waves respectively. The equations for the eigen modes take the from:

$$\begin{pmatrix} [\omega^2 + \alpha^2 C_T^2 - q^2 C_L^2] & \alpha q_{SL}(C_L^2 - C_T^2) & \sim \mu \\ -\alpha q_{SL}(C_L^2 - C_T^2) & [\omega^2 + \alpha^2 C_L^2 - q_{SL}^2 C_T^2] & \sim \mu \\ \sim \mu & \sim \mu & \frac{1}{2}[\omega^2 \rho + \alpha^2 C_L^2] \end{pmatrix} \begin{pmatrix} u_S \\ u_C \\ u_0 \end{pmatrix} = \begin{pmatrix} 0 \\ 0 \\ 0 \end{pmatrix}, \tag{S4}$$

where the small parameter $\mu$ is defined as $\mu \sim \Delta\rho/\langle\rho\rangle \sim \Delta C_L/\langle C_L\rangle \sim \Delta C_T/\langle C_T\rangle \ll 1$. In the limit of infinitely small, i.e., disappearing modulation of the sample parameters, we assume $\mu = 0$. Then, the compatibility condition for Eq. (S4) predicts three bulk acoustic modes differing by their depth propagation constants $\alpha_{z,L,T}$: laterally unmodulated bulk longitudinal $k = 0$ mode ($\alpha_z^2 = -(\omega/C_L)^2 \equiv -k_L^2$), laterally modulated bulk longitudinal $k = 0$ mode ($\alpha_L^2 = q_{SL}^2 - (\omega/C_L)^2 \equiv q_{SL}^2 - k_L^2$) and laterally modulated bulk transversal $k = 0$ mode ($\alpha_T^2 = q_{SL}^2 - (\omega/C_T)^2 \equiv q_{SL}^2 - k_T^2$). Saying differently, the theory predicts a single wave (longitudinal) in the zeroth diffraction order ($m = 0$) and two waves (longitudinal and transverse) waves in the first diffraction order ($m = \pm 1$). Note that the transverse acoustic mode is absent in the zeroth diffraction order via symmetry considerations, because we have limited our analysis to the symmetric modes only. In the presence of a weak modulation of the material parameters in the SL, $\mu \ll 1 \neq 0$, the revealed three modes become coupled because of the scattering on the SL spatial periodicity. However, this scattering leads only to small modifications of the depth propagation constants, i.e., $\sim \mu^2$: $\alpha_{z,L,T} \Rightarrow \alpha_{z,L,T}(1 - \beta_{z,L,T}\mu^2) \equiv \alpha_{z,L,T}^{(\mu)}$. Here $\beta_{z,L,T}$ is of the order $\mu^0$.

Each of the above-derived eigen modes includes the components $u_S$, $u_c$ and $u_0$ (in a particular proportions), and, as a consequence the components of the mechanical displacement include, in general, the contributions from all three eigen modes:

$$u_1 = \sum_{j=z,L,T} 2A_j \sin(q_{SL}x_1) e^{-\alpha_j^{(\mu)}x_3}, \quad u_3 = \sum_{j=z,L,T} A_j \left[\left(\frac{u_0}{u_S}\right)_j + 2\left(\frac{u_C}{u_S}\right)_j \cos(q_{SL}x_1)\right] e^{-\alpha_j^{(\mu)}x_3}. \quad (S5)$$

In the considered case of the weak modulation, $\mu \ll 1$, the proportions of the different components in the displacements structure are controlled in the leading order by the following relations: $\left(\frac{u_C}{u_S}\right)_{z,L,T} \sim \mu^0$, $\left(\frac{u_0}{u_S}\right)_{L,T} \sim \mu$, $\left(\frac{u_0}{u_S}\right)_z \sim \mu^{-1}$. The choice of the amplitudes $A_j$ in the displacements fields Eq. (S5) provides opportunity to satisfy the boundary conditions at mechanically free surface. Substitution of the displacement fields Eq. (S5) and the spatial distributions of the elastic moduli $C_{11}(x_1)$, $C_{12}(x_1)$ and $C_{44}(x_1)$, in the boundary and separation of the constant terms and the terms proportional to $\sin(q_{SL}x_1)$ and to $\cos(q_{SL}x_1)$, results in the system of equations for the displacement amplitudes $A_z, A_L$ and $A_T$:

$$\begin{pmatrix} \gamma_{11} + \delta_{11}\mu^2 & \gamma_{12} + \delta_{12}\mu^2 & \gamma_{13} + \delta_{13}\mu^2 \\ \gamma_{21} + \delta_{21}\mu^2 & \gamma_{22} + \delta_{22}\mu^2 & \gamma_{23} + \delta_{23}\mu^2 \\ \gamma_{31}\mu & \gamma_{32}\mu & \gamma_{33}\mu^{-1} + \delta_{33}\mu \end{pmatrix} \begin{pmatrix} A_L \\ A_T \\ A_z \end{pmatrix} = \begin{pmatrix} 0 \\ 0 \\ 0 \end{pmatrix}. \quad (S6)$$

In Eq. (S6) $\gamma_{ij}$ and $\delta_{ij}$ are the constants of the $\mu^0$ order, and only the terms of the first two leading orders are kept in each of the equations composing the system Eq. (S6). In accordance with Eq. (S6), when $\mu=0$ the laterally unmodulated mode, i.e. $z$ mode, does not contribute to the SAW ($A_z=0$), while the modes $A_L$ and $A_T$ are coupled via: $\begin{pmatrix} \gamma_{11} & \gamma_{12} \\ \gamma_{21} & \gamma_{22} \end{pmatrix}\begin{pmatrix} A_L \\ A_T \end{pmatrix} = \begin{pmatrix} 0 \\ 0 \end{pmatrix}$. Here the matrix $\begin{pmatrix} \gamma_{11} & \gamma_{12} \\ \gamma_{21} & \gamma_{22} \end{pmatrix}$ is proportional to the Rayleigh matrix $\begin{pmatrix} 2\alpha_L & \alpha_T + q_{SL}^2/\alpha_T \\ (c_L^2 - 2c_T^2)q_{SL} - c_L^2\alpha_L^2/q_{SL} & -2c_T^2 q_{SL} \end{pmatrix}$. The Rayleigh determinant is zero, $(2q_{SL}^2 - k_T^2)^2 - 4q_{SL}^2 \alpha_L \alpha_T = 0$, at the frequencies $\omega_{gR}$ of the Rayleigh SAW eigenmode. This solution, obtained in the limit $\mu = 0$, predicts that the Rayleigh gR wave at $\omega = \omega_{gR}$ with the propagation vector equal to $k = q_{SL} \equiv 2\pi/d_{SL}$, i.e., with the initially expected periodicity, belongs in the scheme of folded Brillouin zone to the optical-type branch of the gR wave, $\omega(k = 0) = \omega_{gR}$ (Fig. S2). It propagates at Rayleigh velocity $c_R$ of an « average » medium.

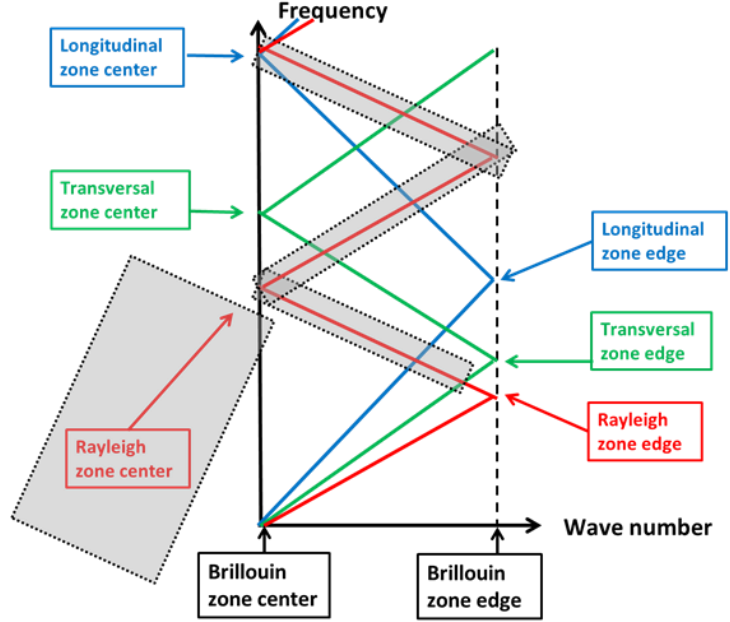

**Figure S2.** Schematic presentation of the surface Rayleigh modes (red lines) and surface skimming transverse and longitudinal bulk modes (green and blue lines, respectively) in the mini Brillouin zone, obtained by formal folding the dispersion relations of the surface and bulk modes in an "average" homogenous medium. The lowest (acoustical) branches of the surface skimming bulk waves describe the boundaries for the existence of the bulk modes. The Rayleigh SAW $\omega(k = 0) = \omega_{gR}$ obtained under the assumptions in Eq. (S3) in the limit $\mu = 0$ is marked as Rayleigh zone center mode. In the limit $\mu \neq 0, \mu \ll 1$, the SAW weakly interacts with the bulk acoustic modes. The parts of the SAW spectrum that are expected to be modified by these interactions are symbolically shaded in grey. The developed theory predicts that this interaction results in small shift of the frequency and in weak attenuation of the gR wave in the Brillouin zone center, i.e., $\Delta Re(\omega_{gR})/\omega_R \sim \mu^2, Im(\omega_{gR})/\omega_{gR} \sim \mu^2$. The revealed gR wave in the limit $\mu \neq 0, \mu \ll 1$ is, in general, a three-component wave, while in the limit $\mu = 0$ it is a two-component wave. Additionally, $\mu \neq 0$ generally results in frequencies splitting of symmetrical and anti-symmetrical gR waves in the center and the edge of the Brillouin zone (not presented in the scheme, but evidenced in Fig. 2 (c) of the manuscript).

When $\mu \neq 0$ all three amplitudes of the bulk modes $A_L, A_T$ and $A_z$ in Eq. (S5) are coupled in the gR wave, which contains a laterally unmodulated part, although smaller than the modulated ones:

$$u_1 \sim \mu^0 sin(q_{SL} x_1), u_3 \sim \mu + \mu^0 cos(q_{SL} x_1). \tag{S7}$$

Equation (S7) describes the lateral structure of the horizontal and vertical components of the mechanical displacement on the free surface. The complete description of the spatial structure of the generalized Rayleigh SAWs in the limit $\mu \neq 0, \mu \ll 1$ could be obtained from Eq. (S6), which predicts that in this limit $A_z \sim \mu^2 A_L \sim \mu^2 A_T$. Therefore, if the corrections $\sim \mu^2$ are omitted, in the description of the gR wave, the spatial structure of the gR wave can be described by

$$u_1 = \sum_{j=L,T} 2A_j \sin(q_{SL} x_1) e^{-\alpha_j x_3}, u_3 = \sum_{j=L,T} A_j \left[ \left( \frac{u_0}{u_S} \right)_j + 2 \left( \frac{u_C}{u_S} \right)_j \cos(q_{SL} x_1) \right] e^{-\alpha_j x_3}, \tag{S8}$$

where the ratio of $A_L$ and $A_T$ can be evaluated in the limit $\mu = 0$. Thus, if the corrections $\sim \mu^2$ are omitted, then the generalized Rayleigh wave is a two-component wave similar to the Rayleigh wave in the limit $\mu = 0$, however the two evanescent bulk components (modes) of the gR wave contain, in comparison with the Ra leigh wave in averaged medium, the laterall unmodulated contributions, i.e., Eu. (S8). In comparison with the frequency of the Raleigh wave predicted in the absence of the parameters modulation, $\omega_{gR} = q_{SL}c_R = \frac{2\pi c_R}{d_{SL}}$, the frequency evaluated with parameters modulation shifts b $\Delta Re(\omega_{gR})/\omega_{gR} \sim \mu^2$ and it acquires an imaginary part $Im(\omega_{gR})/\omega_{gR} \sim \mu^2$.

The above-presented simplest theory, predicts that surface confined modes on a semi-infinite SL stratified normal to its surface contain laterally homogeneous components in their mode structure. This means that, potentially, these modes could be excited even by a laterally homogeneously (averaged over the SL period) part of the distributed photo-induced stresses. This also suggests that they could be detected by normally incident probe radiation as oscillating motion of the laterally homogeneous (averaged over the SL period) near-surface layer of the sample. Importantly, in the case of weak modulation of the materials parameters in the SL, these generalized Rayleigh-type SAWs exhibit weak attenuation.

# Optical detection of generalized Rayleigh-type surface acoustic wave on a superlattice
# stratified normal to its surface

In the picosecond laser ultrasonics, the acoustic waves in the GHz-THz frequency range are detected via the interference on the photodetector of the probe light scattered by them and the probe light scattered by the stationary surfaces/interfaces of the sample. The general approach to predict the transient acoustically induced optical reflectivity changes, monitored by the photodetector, consists of the three following stages. Firstly, the probe light fields reflected by the sample and transmitted into the sample in the absence of the coherent acoustic waves should be evaluated. The probe light reflected by the sample and incident on the photodetector should be predicted. Secondly, the probe light scattered by the acoustic wave either at the sample surface or in the bulk of the sample, or both should be evaluated. The scattered light reaching the photodetector should be predicted. Thirdly, the component of the probe light intensity incident on the photodetector, which is proportional to the product of the amplitudes of the reflected and scattered light should be found. The components of the intensity, which depend only on the intensity of the reflected light, do not provide information on the acoustic waves, while the intensity components which depend only on the intensity of the acoustically scattered light are negligibly small, because of the small amplitude of the acoustically-scattered probe light, in comparison with the amplitudes of the reflected probe light.

To get insightful predictions on the detection of the generalized Rayleigh-type SAWs, which are denoted in the main text of the manuscript as gR waves, in the SL presented in Figs. S1 and S3, we retain all the simplifying assumptions predicting the gR SAWs, leading to their description in Eq. S8, and we introduce both specific and similar simplifying assumptions for the description of the probe light field.

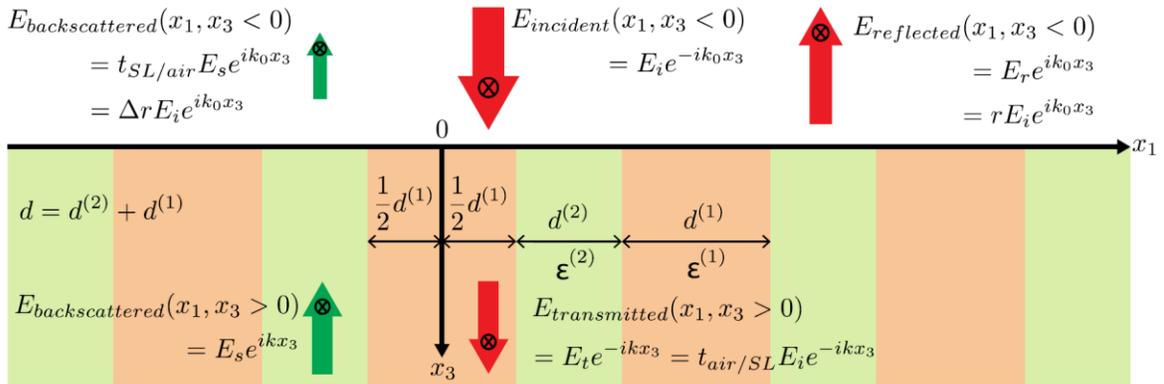

**Figure S3.** Schematic presentation of the time-domain Brillouin scattering of probe light by the generalized Rayleigh SAW in the SL stratified normal to the mechanically free surface. In the considered case of sub-optical superlattices only plane laterally unmodulated reflected, transmitted and scattered waves can be accounted for the description of the leading order optical and acousto-optical processes. $t_{SL/air}$ and $t_{air/SL}$ denote the transmission coefficient of light from the superlattice into the air and in the opposite direction, respectively. The acoustically induced changes in transient optical reflectivity monitored by the photodetector result from the heterodyning at the photodetector of the acoustically backscattered probe light $\sim \Delta r E_i$ by the probe light reflected at the air/SL interface $\sim r E_i$.

The first specific assumption, which makes the problem significantly less cumbersome, while retaining the main features of the detection process, is the postulation that the normally incident probe light field is polarized parallel to the surfaces of the individual layers, i.e., in the ($x_2,x_3$) plane. Under this assumption, although the layered half-space under consideration is, in general, an optically anisotropic/birefringent medium [S6], the probe light incident along the optical symmetry axis, does not exhibit any mode conversion in the reflection/transmission by the SL, and can be described by a single component of the electric field, i.e., $E_{incident} = E_2 \equiv E_i$. The reflected probe light incident on the photodetector, which provides heterodyning of the acoustically scattered light, is the light polarized as the incident one. Therefore, if the probe light is mode converted in its scattering by the acoustic field, then the second process of its mode conversion would be required to allow its interference with the reflected light. This second mode conversion could be also only due to the acoustic waves. Each mode conversion introduces additional smallness in the resulting light field, because the SAWs are of small amplitude. Thus, we omit here the contributions to the transient reflectivity signal of the probe light mode conversion processes. However, it is worth noting that a modified optical scheme, i.e., of the depolarized transient reflectivity measurements, can be suggested for the dedicated study of these processes and the detection of SAWs via these processes, if required.

The probe light reflection/transmission by the SL is described by the following Helmholtz equation

$$\left(\frac{\partial^2}{\partial x_1^2} + \frac{\partial^2}{\partial x_3^2}\right)E + k_0^2 \varepsilon(x_1)E = 0. \tag{S9}$$

Here $k_0$ is the probe wave number in vacuum, $\varepsilon(x_1,x_3>0)$ is the relative permittivity of the SL, while the permittivity of air is $\varepsilon(x_1,x_3<0)\equiv 1$. Under the assumptions identical to those leading to Eq. (S3), the distributions of the relative permittivity in the SL and the probe light eigen modes in the system presented in Fig. S3 have the following simplest descriptions

$$\varepsilon(x_1, x_3 > 0) = \langle\varepsilon\rangle + 2\Delta\varepsilon \cos(q_{SL}x_1), \langle\varepsilon(x_3 > 0)\rangle = \frac{d^{(1)}\varepsilon^{(1)} + d^{(2)}\varepsilon^{(2)}}{d^{(1)} + d^{(2)}},$$

$$\Delta\varepsilon(x_3 > 0) = = \frac{1}{\pi}\left(\varepsilon^{(1)} - \varepsilon^{(2)}\right)\cos\left[\frac{\pi}{2}\frac{|d^{(1)} - d^{(2)}|}{(d^{(1)} + d^{(2)})}\right],$$

$$E(x_1, x_3) = [\langle E\rangle + 2E_C\cos(q_{SL}x_1)]e^{-\beta x_3}. \tag{S10}$$

Here $\beta$ denotes the propagation/penetration constant, related to the projection of the optical wave vector on the $x_3$ direction. The approximation suggested in Eq. (S10), results in the following solution of Eq. (S9) for the reflected probe light,

$$E_{reflected}(x_1, x_3 < 0) = \langle E_r\rangle e^{ik_0 x_3} + 2E_{r,C}\cos(q_{SL}x_1)e^{i\sqrt{k_0^2 - q_{SL}^2}x_3}. \tag{S11}$$

Note, that the second contribution to the reflected light in Eq. (S11) is an evanescent wave for the considered-by-us sub-optical SLs, as $k_0 < q_{SL}$. Thus only the plane (laterally unmodulated) component of the reflected probe light reaches the photodetector for the interference with the acoustically scattered probe light and, consequently, only the plane components of the scattered light reaching the photodetector could contribute to the transient reflectivity signal. The substitution of Eq. (S10) into Eq. (S9) leads to the description of the optical eigen modes inside the SL.

$$\begin{pmatrix} \beta^2 - q_{SL}^2 + k^2 & \left(\frac{\Delta\varepsilon}{\langle\varepsilon\rangle}\right)k^2 \\ 2\left(\frac{\Delta\varepsilon}{\langle\varepsilon\rangle}\right)k^2 & \beta^2 + k^2 \end{pmatrix}\begin{pmatrix} E_C \\ \langle E\rangle \end{pmatrix} = \begin{pmatrix} 0 \\ 0 \end{pmatrix}, \quad k^2 \equiv \langle\varepsilon\rangle k_0^2 \equiv n^2 k_0^2. \tag{S12}$$

Equation (S12) predicts inside the SL two optical modes with the different propagation/penetration constants,

$\beta_\mp^2 = -k^2 + \frac{q_{SL}^2}{2}\left[1 \mp \sqrt{1 + 2\left(2\frac{k^2}{q_{SL}^2}\right)^2\left(\frac{\Delta\varepsilon}{\langle\varepsilon\rangle}\right)^2}\right] \equiv -k^2 + \frac{q_{SL}^2}{2}\left[1 \mp \sqrt{1 + 2\eta^2}\right]$ Here a compact notation $\eta$ is introduced for a parameter $\eta \equiv \left(2\frac{k^2}{q_{SL}^2}\right)\left(\frac{\Delta\varepsilon}{\langle\varepsilon\rangle}\right) = \left(2\frac{d_{SL}^2}{\lambda^2}\right)\left(\frac{\Delta\varepsilon}{\langle\varepsilon\rangle}\right) \ll 1$, which is small in sub-optical SLs with $k^2 \ll q_{SL}^2$ even in the case of strong optical contrast between the individual SL layers, i.e., $\Delta\varepsilon \sim \langle\varepsilon\rangle$. The existence of a specific small parameter $\eta$ is another specific assumption in the theory of the acousto-optic detection of SAWs in comparison with the theory of the SAW eigen modes in Supplementary S1. Neglecting the corrections of the order $\eta^2$ to the propagation constants and to the modes structure in general, the transmitted and acoustically backscattered probe light fields inside the SL take the following forms

$$E_{transmitted}(x_1, x_3 > 0) = A_t^q[-\eta + 2\cos(q_{SL}x_1)]e^{-qx_3} + A_t^{ik}[1 + \eta\cos(q_{SL}x_1)]e^{-ikx_3}, \tag{S13}$$

$$E_{backscattered}(x_1, x_3) = A_s^{-q}[-\eta + 2\cos(q_{SL}x_1)]e^{qx_3} + A_s^{-ik}[1 + \eta\cos(q_{SL}x_1)]e^{ikx_3}. \tag{S14}$$

In the single scattering approximation, valid due to the smallness of the acoustic strains, it is possible, first, to neglect acoustically scattered field, and, from the conditions of the continuity at the interface of the electrical field component $E_2$ and of the magnetic field component $H_1 \sim \partial E_2/\partial x_3$, to find the reflection/transmission coefficients, defining the amplitudes of the reflected transmitted probe fields

$$r \equiv \frac{\langle E_r\rangle}{E_i} \cong -\frac{n-1}{n+1}, \quad t_{\frac{air}{SL}} \equiv \frac{A_t^{ik}}{E_i} \cong \frac{2}{n+1}, \quad \frac{A_t^q}{E_i} \cong -\frac{\eta}{n+1}\frac{\left(n + \sqrt{1 - \frac{\lambda_0^2}{a^2}}\right)}{\left(\sqrt{n^2 - \frac{\lambda_0^2}{a^2}} + \sqrt{1 - \frac{\lambda_0^2}{a^2}}\right)} \approx -\frac{1}{2(n+1)}\eta \ll 1. \tag{S15}$$

The solutions in Eq. (S15), where only the leading order terms in the powers of the small parameter $\eta \ll 1$ has been retained, indicate the smallness of the probe optical field transmitted in the first diffraction order and the direct relation of this smallness to the sub-optical periodicity of the SL.

The weakness of the coupling between the different diffraction orders (when light is transmitted across the air/SL interface in the considered sub-optical SL) revealed by Eq. (S15) suggests that, if the detection process is possible in the asymptotic case $\eta = 0$, then accounting for the corrections of its description in the case $\eta \neq 0, \eta \ll 1$ would be unnecessary. In the limit $\eta = 0$, the transmitted light field in Eq. (S13) reduces to $E_{transmitted}(x_1, x_3 > 0, \eta = 0) = A_t^{ik}e^{-ikx_3}$ and only the plane unmodulated acoustically backscattered field, i.e., in the zeroth diffraction order in Eq. (S14), can contribute (after the transmission from the SL into air) to signal of our interest, $E_{backscattered}(x_1, x_3 > 0, \eta = 0) = A_s^{-ik}e^{ikx_3}$. These fields are coupled by the acoustic field in the SLs in the following Helmholtz equation

$$\left(\frac{\partial^2}{\partial x_1^2} + \frac{\partial^2}{\partial x_3^2}\right)E + k_0^2[\varepsilon(x_1) + \varepsilon_{acoust}(x_1, x_3)]E = 0,$$

where $\varepsilon_{acoust}(x_1, x_3)$ denotes acoustically induced changes in the permittivity. In the single scattering approximation, this equation describes the emission of the scattered probe light by the acoustically-induced non-linear polarization

$$\frac{\partial^2}{\partial x_3^2}E_s + k_0^2\langle\varepsilon\rangle E_s = -k_0^2\langle\varepsilon_{acoust}(x_3)\rangle E_{transmitted}. \tag{S16}$$

Here the lateral averaging over the period of the SL takes into account the fact that the plane (laterally unmodulated) transmitted probe light field $E_{transmitted}$ can be scattered into the plane (laterally unmodulated) light only by the laterally unmodulated (averaged) component of the acoustically induced changes of permittivity. Note that Eq. (S16) accounts for both forward and backward propagating scattered light but only the evaluation of the backward scattered probe light incident on SL/air interface is required for the evaluation of the signal. The required solution of Eq. (S16) leads, finally, to the following presentation for the amplitude of the acoustically scattered probe light contributing to the changes in the transient optical reflectivity monitored by the photodetector (Fig. S3)

$$\Delta r = -\frac{ik}{2}(1-r^2)\int_0^\infty \frac{\langle\varepsilon_{acoust}(x_3)\rangle}{\langle\varepsilon\rangle}e^{-2ikx_3}dx_3. \tag{S17}$$

If we introduce the notation $\delta f$ for the acoustically induced variation of the physical parameter $f$ and assume that the SLs for the generation and the detection of SAW are identical (as it is in our experiments), then, in the gR SAW, Eq. (S8), acting on the electrical permittivity, Eq. (S10), $\langle\varepsilon_{acoust}(x_1,x_3)\rangle$ can be written as

$$\langle\varepsilon_{acoust}(x_1, x_3)\rangle \equiv \langle\delta\varepsilon(x_1, x_3)\rangle = \langle\delta\langle\varepsilon\rangle + 2\delta(\Delta\varepsilon)\cos(q_{SL}x_1)\rangle.$$

The first term in this expression describes the possibility to detect generalized Rayleigh SAWs due to the fact that its spatially unmodulated component modulates in time the spatially-averaged part of the SL electrical permittivity. The second term predicts the possibility to detect generalized Rayleigh SAWs due to the interaction (parametric) of the spatially modulated ($\sim\cos(q_{SL}x_1)$) component of SAWs with the spatially modulated part of the electrical permittivity.

An intermediate result, which is obtained with the use of Eq. (S10) for the SL with the equal thicknesses of two layers, $d^{(1)}=d^{(2)}$ reads

$$\langle\varepsilon_{acoust}(x_1,x_3)\rangle = \left(\varepsilon^{(1)} - \varepsilon^{(2)}\right)\frac{\delta d^{(1)} - \delta d^{(2)}}{2a} + \langle\frac{\delta\varepsilon^{(1)} + \delta\varepsilon^{(2)}}{2} + 2\frac{\delta\varepsilon^{(1)} - \delta\varepsilon^{(2)}}{\pi}\cos(q_{SL}x_1)\rangle. \tag{S18}$$

Here the terms $\sim(\delta d^{(1)}-\delta d^{(2)})$ and $\sim(\delta\varepsilon^{(1)}+\delta\varepsilon^{(2)})$ are contributions from $\langle\delta\langle\varepsilon\rangle\rangle$. Equation (S18) attracts our attention to the fact that acousto-optic detection of SAWs, even in the SLs with constant period, which is not modified by the symmetric generalized Rayleigh SAW under consideration here and $\delta d^{(1)}+\delta d^{(2)}=0$, could take place not only due to the photo-elastic effect in the individual layers. If $\varepsilon^{(1)}\neq\varepsilon^{(2)}$, the detection can be also due to the geometrical effect of the acoustically induced variations in the thicknesses of the layers. For the SAW in Eq. (S8), this geometrical contribution can be evaluated as follows

$$\frac{\langle \varepsilon_{acoust}(x_3)\rangle_{geom}}{(\varepsilon^{(1)}-\varepsilon^{(2)})} = \frac{d^{(1)}}{d_{SL}} = \frac{u_1(x_1=d_{SL}/4, x_3)-u_1(x_1=-d_{SL}/4, x_3)}{a} = \frac{2u_1(x_1=\frac{d_{SL}}{4},x_3)}{a} = \frac{4}{d_{SL}}\sum_{j=L,T} A_j e^{-\alpha_j x_3}. \quad (S19)$$

The derivation of Eq. (S19) assumes that only the horizontal components of the mechanical displacement in SAWs contribute to the modification (breathing) of the layers thicknesses. The contributions into $\langle \varepsilon_{acoust}(x_3)\rangle$ from the photo-elastic effect in the individual layers are described by

$$\delta\varepsilon^{(i)} = -(\varepsilon^{(i)})^2 p_{21}^{(i)}\left(\frac{\partial u_1}{\partial x_1}+\frac{\partial u_3}{\partial x_3}\right) \equiv \bar{p}^{(i)}\left(\frac{\partial u_1}{\partial x_1}+\frac{\partial u_3}{\partial x_3}\right)$$
$$= \bar{p}^{(i)} \sum_{j=L,T} A_j \left\{-\alpha_j \left(\frac{u_0}{u_S}\right)_j + 2\left[q - \alpha_j\left(\frac{u_C}{u_S}\right)_j\right] \cos(q_{SL}x_1)\right\} e^{-\alpha_j x_3}.$$

Here $p_{21}^{(i)}$ denote the relevant component of the photo-elastic tensor [S7]. Substitution of this result into Eq. (S18) and averaging over the SL period leads to the description of photo-elastic contribution into $\langle \varepsilon_{acoust}(x_1, x_3)\rangle$

$$\langle \varepsilon_{acoust}(x_3)\rangle_{photo-elastic} = \sum_{j=L,T} A_j \left\{-\alpha_j\left(\frac{u_0}{u_S}\right)_j \frac{\bar{p}^{(1)}+\bar{p}^{(2)}}{2} + 2\left[q-\alpha_j\left(\frac{u_C}{u_S}\right)_j\right]\frac{\bar{p}^{(1)}-\bar{p}^{(2)}}{\pi}\right\} e^{-\alpha_j x_3}. \quad (S20)$$

The first combination in the figure brackets of (S20) predicts the detection due to the averaged photo-elasticity of the SL, $\frac{\bar{p}^{(1)}+\bar{p}^{(2)}}{2}$, of the laterally unmodulated (averaged) component of the generalized Rayleigh wave, $\sim\left(\frac{u_0}{u_S}\right)_j \sim \mu$ (see Eqs. (S7) and (S8)). The second combination in the figure brackets of (S20), which is due the averaging of the product of the laterally modulated components of the SAW and laterally modulated component of the SL photo-elasticity. It appears due to $\langle \cos^2(q_{SL}x_1)\rangle = 1/2 \neq 0$ and predicts the changes in the averaged acousto-optical response of the SL due to the coupling of the SL periodicity to the laterally modulated components of the SAW. It is worth mentioning here that we call Eq. (S20) the photo-elastic contribution, because it would disappear with diminishing components of the photo-elastic tensor $p_{21}^{(i)}$, however the parameter controlling this contribution, i.e., $\bar{p}^{(i)}$, is additionally proportional to the square of the permittivity, $\bar{p}^{(i)} \equiv -(\varepsilon^{(i)})^2 p_{21}^{(i)}$.

The results in Eqs. (S19) and (S20) demonstrate that even when $d_{SL}/\lambda \to 0$, i.e., in deeply sub-optical SLs, the detection of deeply sub-optical generalized Rayleigh SAWs is possible. It is theoretically possible even in the absence of the contrast between both the permittivity and photo-elastic constants of the individual layers. In this case, the SAW is detected via its laterally unmodulated component, which is an essential feature of the generalized Rayleigh waves in the considered nanostructured sample. However, in the case of weak contrast between both the densities and the elastic moduli of the individual layers, i.e., in the regime $\mu \ll 1$ evaluated in details in Supplementary 1, the dominant contributions to the detection of SAWs are expected due to the contrast in either permittivity, or photo-elastic constant, or both, of the individual layers. This contrast has not been assumed a weak one in the above-presented theoretical approach. Moreover, for the effective generation of SAWs by the pump laser pulses, the optical contrast between the individual layers is profitable to increase, e. g., combining the layers opaque and transparent to pump light. Thus, in the limiting case, where both $\mu \to 0$ and $d_{SL}/\lambda \to 0$, the acoustically-induced changes in electric field reflectivity are described by

$$\frac{\Delta r}{r} = \frac{4ik}{(\varepsilon^{(1)}+\varepsilon^{(2)})}\left(r - \frac{1}{r}\right)\sum_{j=L,T}\frac{A_j}{\alpha_j+2ik}\left\{\left(\varepsilon^{(1)} - \varepsilon^{(2)}\right) + \left[1 - \frac{d_{SL}\alpha_j}{2\pi}\left(\frac{u_C}{u_S}\right)_j\right]\left(\bar{p}^{(1)} - \bar{p}^{(2)}\right)\right\}. \quad (S21)$$

This solution is obtained by substitution of Eqs. (S19) and (S20) in Eq. (S17). If we take into account that the penetration depths of the gR SAW components are of the order of the SAW wavelength, i.e., $\alpha_j \sim 1/d_{SL}$, then, in the considered limit $d_{SL}/\lambda \to 0$, Eq. (S21) additionally simplifies to

$$\frac{\Delta r}{r} = \frac{4ik}{(\varepsilon^{(1)}+\varepsilon^{(2)})}\left(r - \frac{1}{r}\right)\sum_{j=L,T}\frac{A_j}{\alpha_j}\left\{\left(\varepsilon^{(1)} - \varepsilon^{(2)}\right) + \left[1 - \frac{d_{SL}\alpha_j}{2\pi}\left(\frac{u_C}{u_S}\right)_j\right]\left(\bar{p}^{(1)} - \bar{p}^{(2)}\right)\right\}, \quad (S22)$$

demonstrating that $\Delta r/r \sim d_{SL}/\lambda$. Thus, the SAW induced transient reflectivity signal inevitably diminishes with the diminishing pitch of the SL even in its leading order retained in Eq. (S22). However, this diminishing is much slower than in some other processes of light scattering by sub-optical inhomogeneities/objects. For example, the Rayleigh scattering of light by sub-optical spheres scales $\sim (d_{SL}/\lambda)^4$ [S8]. Moreover, it is important to mention here that to reach the regime, which we denote for the simplicity of the above discussions as $d_{SL}/\lambda \to 0$, could be not that easy under some practical circumstances. In fact, the precise condition for the transition from Eq. (S21) to Eq. (S22) is not $d_{SL}/\lambda \ll 1$, but $4\pi|n|d_{SL}/\lambda_0 \ll 1$, where $|n|$ denotes the modulus of the averaged refractive index of the SL for the probe light of the wavelength $\lambda_0$ in vacuum. In the considered SL with the equal thicknesses of the individual layers, even if one of the two different layers strongly absorbs or reflects probe light, then $\frac{\lambda_0}{4\pi|n|} \approx 20 - 30$ nm. In this case, the asymptotic $\Delta r/r \sim d_{SL}/\lambda$ could be reached only in the SLs with the single digit nanometers period.

To conclude it is worth reminding here that in our optical scheme only the amplitude of $\Delta r/r$ and not its phase is monitored. The signal in this optical scheme is due to the changes in the light intensity reflectivity, $R \equiv rr^*$, $dR/R \equiv 2Re(\Delta r/r)$ [S8], where "*" denotes the complex conjugation. This is one of the reasons for omitting in the above presented theory the contribution to the detection process of SAWs from the motion of the air/SL interface induced by them [S9]. In fact, in analogy with the above presented theoretical analysis, one could expect that the generalized Rayleigh SAW in addition to laterally surface ripples, could induce the average surface displacement. However, this averaged surface displacement modifies only the phase and not the amplitude of $\Delta r/r$. This averaged surface motion could be monitored by the optical interferometry or by the beam deflection technique, if required. At the same time the SAW induced ripples with the sub-optical periodicity are diffracting the incident probe light into the evanescent orders (see the second term in Eq. (S11)), where it does not reach the photodetector.